\documentclass[onecolumn,amsmath,amssymb,aps]{revtex4-2}
\usepackage{graphicx}% Include figure files
\usepackage{dcolumn}% Align table columns on decimal point
\usepackage{bm}% bold math
\usepackage{hyperref}% add hypertext capabilities
\usepackage{tikz}
\usetikzlibrary{arrows.meta}
\usepackage{booktabs}

\newtheorem{proposition}{Proposition}

\begin{document}

\preprint{APS/123-QED}

\title{Revisitation of algebraic approach for time delay interferometry}% Force line breaks with \\

\author{Weisheng Huang}
\author{Pan-Pan Wang}\email[E-mail: ]{ppwang@hust.edu.cn}
\author{Yu-Jie Tan}
\author{Cheng-Gang Shao}\email[E-mail: ]{cgshao@hust.edu.cn}

\affiliation{MOE Key Laboratory of Fundamental Physical Quantities Measurement, Hubei Key Laboratory of Gravitation and Quantum Physics, PGMF, and School of Physics, Huazhong University of Science and Technology, Wuhan 430074, China}

\date{\today}% It is always \today, today,
             %  but any date may be explicitly specified

\begin{abstract}
    Time Delay Interferometry (TDI) is often utilized in the data pre-processing of space-based gravitational wave detectors, primarily for suppressing laser frequency noise. About twenty years ago, assuming armlengths remain constant over time, researchers presented comprehensive mathematical descriptions for the first-generation and modified first-generation TDI. However, maintaining a steady distance between satellites is pragmatically challenging. Hence, the operator equation that neutralizes laser frequency noise, though provided, was deemed difficult to resolve. In this paper, we solve this equation in the context of a non-static scenario where distances between spacecrafts vary over time. Surprisingly, contrary to what previous researchers thought, the study reveals that the equation has only the zero solution, which suggests that no nonzero TDI combination can entirely suppress laser frequency noise under time-varying armlengths. This necessitates the persistent search for second-generation TDI combinations through alternative methods besides directly solving the operator equation. We establish the connections between TDI combinations of different generations and propose a search strategy for finding higher-generation TDI combinations by using generators of lower-generation TDI. The findings contribute to the ongoing discussion on gravitational waves and provide a novel insight into the hurdles faced in space-based gravitational wave detection.
\end{abstract}

%\keywords{Suggested keywords}%Use showkeys class option if keyword
                              %display desired
\maketitle

%\tableofcontents

\section{Introduction}\label{sec:1}

Gravitational waves, ripples in spacetime curvature triggered by violent transformations of matter and energy in astronomical phenomena such as the Big Bang and black hole mergers, have been the focus of international scientific inquiry in recent years. A century ago, Einstein predicted their existence in his theory of general relativity, thereby sparking a fascination that persists among scientists worldwide. Distinct from electromagnetic waves, gravitational waves offer a new window through which we can observe the universe. They provide a novel pathway and tool for human exploration and understanding of the cosmos.

In the late 1980s and early 1990s, numerous countries initiated research in this domain. Around ten ground-based gravitational wave detection projects have been proposed internationally, but due to technical challenges and high costs, only a handful persist. In 2016, a milestone was reached when the United States' ground-based LIGO experiment announced the first direct observation of gravitational waves \cite{Abbott2016Observation}, produced by the merger of two black holes in the universe. This marked the inauguration of a new era in gravitational wave astronomy.

Gravitational waves of varying frequencies correspond to different periods and astronomical physical processes in the universe. Space-based detections, unlike ground-based detections, can detect gravitational wave signals in the mid-to-low frequency range~\cite{Pau2017Laser}. This capability allows us to uncover gravitational wave sources with larger masses and greater distances, thus elucidating more complex astronomical phenomena~\cite{Pau2023Astrophysics}. However, given the extremely weak nature of gravitational wave signals, implementing space-based gravitational wave detection demands significant advancements in our current precision measurement and control technologies. These challenges notwithstanding, the pursuit of understanding gravitational waves continues to be a vibrant field of research.

For the frequency band of $10^{-4}-1$ Hz, many space-based gravitational wave detection projects have been proposed, such as LISA~\cite{Pau2017Laser,Pau2023Astrophysics}, TianQin~\cite{Luo2016TianQin} and Taiji~\cite{Hu2017Taiji}, which usually utilize an equilateral triangular formation of three satellites. Unlike ground-based gravitational wave detectors, maintaining a same constant distance between any two satellites in space-based detectors is nearly impossible. Consequently, this causes laser frequency noise in the observed signal with a magnitude of 7-8 orders higher than that of typical gravitational waves. To address this issue, it is necessary to use Time Delay Interferometry (TDI)~\cite{Tinto1999Cancellation,Tinto2021Time,Muratore2022Time}, which suppresses the laser frequency noise below secondary noises before further extraction of the gravitational wave signal.

For the static configuration of non-equal armlengths, the laser frequency noise can be completely cancelled in the observed signals using TDI technique. By utilizing the theory of syzygies on moduli, a complete mathematical description can be provided for the first-generation generation and modified first-generation TDI~\cite{Dhurandhar2002,Rajesh2004}. For the non-static configuration, where the distances between satellites vary over time, although we can obtain an equation
\begin{equation}\label{eq:1-1}
	q_3 (1 - D_2 D_3 D_1) + q_{1'} (D_3 D_1 - D_{2'}) +q_{2'} (D_1 - D_{3'} D_3 D_1) + q_{3'} (1 - D_{1'} D_1) = 0
\end{equation}
regarding the time-delay operator, it is generally considered very difficult to solve~\cite{Dhurandhar2009Time,Tinto2021Time}. Therefore, alternative approaches have been proposed to search for the second-generation TDI combinations, such as Geometric TDI~\cite{Vallisneri2005Geometric,Hartwig2022Characterization,Wang2022Geometric}, Matrix-based approaches~\cite{Vallisneri2021Time,Tinto2021Matrix,Dhurandhar2022Varied}, Bayesian TDI~\cite{Page2021Bayesian}, Combinatorial algebraic approach~\cite{Dhurandhar2010Time,Wu2023Combinatorial} and Lifting procedure~\cite{Tinto2023Second}. In this paper, we solve Eq. \eqref{eq:1-1} and find that it only has a zero solution, which implies that there exists no nonzero TDI combination that can completely suppress laser frequency noise in the context of time-varying armlengths. This result indicates that we can only search for the second-generation TDI combinations through methods other than by directly solving Eq. \eqref{eq:1-1}.

This paper is organized as follows. In section \ref{sec:2}, the configuration of typical space-based gravitational wave detectors and the their associated symbol conventions are introduced. This section also derives operator equation \eqref{eq:1-1}, emphasizing our key finding (Proposition \ref{prop:1}): Eq. \eqref{eq:1-1} possesses only the zero solution. Section \ref{sec:3} focus on a review of the results related to TDI of different generations, namely the zeroth-generation, first-generation, modified first-generation and second-generation TDI. Here, we establish the connections between TDI combinations of different generations. In section \ref{sec:4}, we propose a search strategy for finding higher-generation TDI combinations through generators of lower-generation TDI combinations. Section \ref{sec:5} provides a comprehensive mathematical description of equivalent TDI combinations and outlines inequivalent TDI combinations within generation sets of zeroth-generation TDI, first-generation TDI, and modified first-generation TDI. Section \ref{sec:6} is the conclusions and discussion. Finally, we supply complementary derivations and resources in the appendices. Appendix \ref{appendix:A} includes a detailed proof of Proposition \ref{prop:1}. In Appendix \ref{appendix:B}, we present a general method for solving linear equations over commutative polynomial rings. Appendix \ref{appendix:C} encompasses a discussion on the various orderings about polynomials encountered during the solving process.

\section{Data streams and laser noise cancellation}\label{sec:2}

\begin{figure}[!ht]
	\centering
	\includegraphics[width=0.45\textwidth]{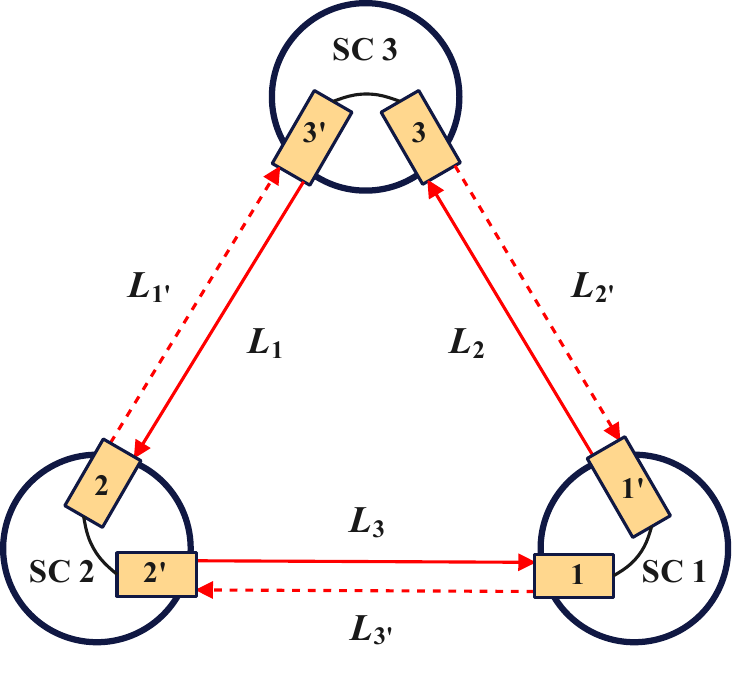}
	\caption{Schematic diagram of the space-based gravitational wave detector.}\label{fig:configuration}
\end{figure}

As shown in Fig. \ref{fig:configuration}, a typical space-based gravitational wave detector consists of three identical spacecrafts, which are labeled SC 1, SC 2, and SC 3 in the clockwise direction. Each spacecraft has two mostly identical optical benches that shoot and recieve lasers. The data streams obtained by laser interferometric measurements among the optical benches include science data streams $s_i$ and $s_{i'}$,  test mass data streams $\varepsilon_i$ and $\varepsilon_{i'}$, reference data streams $\tau_i$ and $\tau_{i'}$, which are defined as follows:
\begin{widetext}
  \begin{equation}\label{eq:2-1}
    \begin{aligned}
      s_i(t) &= h_i(t) + D_{i - 1} p_{(i + 1)'}(t) - p_i(t) + \nu_{(i + 1)'}\left[\vec{n}_{i - 1} \cdot D_{i - 1} \vec{\Delta}_{(i + 1)'}(t) + \vec{n}_{(i - 1)'} \cdot \vec{\Delta}_i(t)\right]  + N_i^S(t), \\
      \varepsilon_i(t) &= p_{i'}(t) - p_i(t) + \mu_{i'}(t) - 2 \nu_{i'}\left[\vec{n}_{(i - 1)'} \cdot \vec{\delta}_i(t) - \vec{n}_{(i - 1)'} \cdot \vec{\Delta}_i(t)\right], \\
      \tau_i(t) &=p_{i'}(t) - p_i(t) + \mu_{i'}(t),
    \end{aligned}
  \end{equation}
\end{widetext}
and
\begin{widetext}
  \begin{equation}\label{eq:2-2}
    \begin{aligned}
      s_{i'}(t) &= h_{i'}(t) + D_{(i + 1)'} p_{i - 1}(t) - p_{i'}(t) + \nu_{i - 1}\left[\vec{n}_{i + 1} \vec{\Delta}_{i'}(t) + \vec{n}_{(i + 1)'} \cdot D_{(i + 1)'} \vec{\Delta}_{i - 1}(t)\right]  + N_{i'}^S(t), \\
      \varepsilon_{i'}(t) &= p_i(t) - p_{i'}(t) + \mu_i(t) - 2 \nu_i \left[\vec{n}_{(i - 1)} \cdot \vec{\delta}_{i'}(t) - \vec{n}_{(i + 1)} \cdot \vec{\Delta}_{i'}(t) \right], \\
      \tau_{i'}(t) &= p_i(t) - p_{i'}(t) + \mu_i(t).
    \end{aligned}
  \end{equation}
\end{widetext}

In the formulas, $h_i$ and $h_{i'}$ are gravitational wave signal, included in the scientific data stream; $p_i$ and $p_{i'}$ are the laser frequency noise, which are 7-8 orders of magnitude larger than the typical gravitational wave signal; $\nu_i$ and $\nu_{i'}$ are the laser frequency; $\vec{n}_i$ and $\vec{n}_{i'}$ are the unit direction vector along the armlengths; $\vec{\Delta}_i$ and $\vec{\Delta}_{i'}$ are the optical bench motion noise; $N_i^S$ and $N_{i'}^S$ are the aggregate optical-path-noises; $\mu_i$ and $\mu_{i'}$ are the fiber noise; $\vec{\delta}_i$ and $\vec{\delta}_{i'}$ are the test mass noise. The time-delay operators $D_i$ are applied to the data steam $f(t)$ as follows:
\begin{equation}\label{eq:2-3}
	\begin{aligned}
		D_i f(t) &= f(t - L_i(t)), \\
		D_j D_i f(t) &= D_j f(t - L_i(t)) \\
		&= f(t - L_j(t) - L_i(t - L_j(t))),
	\end{aligned}
\end{equation}
where $L_i$ and $L_j$ represent armlengths in Fig. \ref{fig:configuration} and the speed of light $c$ is assumed to be 1.

In the paper, we focus on the cancellation of laser frequency noise $p_i$ and $p_{i'}$. By employing various combinations of the 18 available data streams, 6 laser noises $p_i$ and $p_{i'}$ are reduced to 3 laser noises $p_i$~\cite{Tinto2021Time}. The terms containing only the laser frequency noise within the six auxiliary data streams are
\begin{equation}\label{eq:2-4}
	\begin{aligned}
		\eta_i(t) & = D_{i - 1} p_{i + 1}(t) - p_i(t), \\
		\eta_{i'}(t) & = D_{(i + 1)'} p_{i - 1}(t) - p_i(t).
	\end{aligned}
\end{equation}
The TDI combination is denoted by
\begin{equation}\label{eq:2-5}
	\mathrm{TDI} = \sum_{i=1}^3\left(q_i \eta_i + q_{i'} \eta_{i'}\right),
\end{equation}
where $q_i$ and $q_{i'}$ are polynomials of the time-delay operators $D_i$ and $D_{i'}$.

We express $p_i$, $q_i$, $\eta_i$ as column vectors $\mathbf{p}$, $\mathbf{q}$, $\boldsymbol{\eta}$, i.e.,
\begin{equation}\label{eq:2-6}
	\begin{aligned}
		\mathbf{p} &= (p_1, p_2, p_3)^T, \\
		\mathbf{q} &= (q_1, q_2, q_3, q_{1'}, q_{2'}, q_{3'})^T, \\
		\boldsymbol{\eta} &= (\eta_1, \eta_2, \eta_3, \eta_{1'}, \eta_{2'}, \eta_{3'})^T,
	\end{aligned}
\end{equation}
where superscript ``T'' means the transpose.

According to Eq. \eqref{eq:2-4}, one has
\begin{equation}\label{eq:2-7}
	\boldsymbol{\eta} = M_2 \mathbf{p},
\end{equation}
where
\begin{equation}\label{eq:2-8}
	M_2 =
	\begin{pmatrix}
		-1 & D_3 & 0 \\
		0 & -1 & D_1 \\
		D_2 & 0 & -1 \\
		-1 & 0 & D_{2'} \\
		D_{3'} & -1 & 0 \\
		0 & D_{1'} & -1
	\end{pmatrix}.
\end{equation}

In order to cancel the laser frequency noise $\mathbf{p}$, the TDI must satisfy
\begin{equation}\label{eq:2-9}
	\mathrm{TDI} = \mathbf{q}^T \boldsymbol{\eta} = \mathbf{q}^T M_2 \mathbf{p}= 0.
\end{equation}
Thus, we have
\begin{equation}\label{eq:2-10}
	\mathbf{q}^T M_2 = (0, 0, 0),
\end{equation}
which is equivalent to
\begin{equation}\label{eq:2-11}
	\left\{
	\begin{aligned}
		q_1 + q_{1'} - q_{2'} D_{3'} - q_3 D_2 &= 0, \\
		q_2 + q_{2'} - q_{3'} D_{1'} - q_1 D_3 &= 0, \\
		q_3 + q_{3'} - q_{1'} D_{2'} - q_2 D_1 &= 0.
	\end{aligned}
	\right.
\end{equation}
Based on the first two equations, we can obtain $q_1$ and $q_2$ in terms of $q_3$, $q_{1'}$, $q_{2'}$ and $q_{3'}$. Therefore, after substituting $q_1$ and $q_2$ into the third equation in Eq. \eqref{eq:2-11}, we only need to solve the following single equation
\begin{equation}\label{eq:2-12}
	q_3 (1 - D_2 D_3 D_1) + q_{1'} (D_3 D_1 - D_{2'}) +q_{2'} (D_1 - D_{3'} D_3 D_1) + q_{3'} (1 - D_{1'} D_1) = 0.
\end{equation}

Without causing confusion, three different expressions, the TDI satisfying Eq. \eqref{eq:2-9}, the polynomial vector $\mathbf{q}$ satisfying Eq. \eqref{eq:2-10}, and the polynomial vector $(q_1, q_{1'}, q_{2'}, q_{3'})$ satisfying Eq. \eqref{eq:2-11}, are all referred to TDI solutions. Specifically, let
\begin{equation}\label{eq:2-13}
	\mathcal{K} = K \langle D_1, D_2, D_3, D_{1'}, D_{2'}, D_{3'} \rangle
\end{equation}
denote the noncommutative polynomial ring in six variables $D_1, D_2, D_3, D_{1'}, D_{2'}, D_{3'}$ with coefficients in the field $K$. The homomorphism $\varphi: \mathcal{K}^4 \to \mathcal{K}^1$ is defined by
\begin{equation}\label{eq:2-14}
	\varphi(q_3, q_{1'}, q_{2'}, q_{3'}) = q_3 (1 - D_2 D_3 D_1) + q_{1'} (D_3 D_1 - D_{2'}) +q_{2'} (D_1 - D_{3'} D_3 D_1) + q_{3'} (1 - D_{1'} D_1).
\end{equation}
The set of solutions of Eq. \eqref{eq:2-12} is the kerner $\varphi^{-1}(0) \subset \mathcal{K}^4$ of the homomorphism $\varphi$, which is a submodule of $\mathcal{K}^4$~\cite{Tinto2021Time}. 

According to the noncommutativity of the time-delay operators $D_i$ and $D_{i'}$, it is generally considered very difficult to search for solutions of Eq. \eqref{eq:2-11} or Eq. \eqref{eq:2-12} in the literatures~\cite{Rajesh2004,Dhurandhar2009Time,Tinto2021Time}. Surprisingly, contrary to what previous researchers thought, we prove that Eq. \eqref{eq:2-12} has no nonzero solution in the paper.
\begin{proposition}\label{prop:1}
	\textnormal{Eq.} \eqref{eq:2-12} has only zero solution, namely, the kerner of the homomorphism $\varphi$ is
	\begin{equation}
		\varphi^{-1}(0) = \{ (0, 0, 0, 0) \}.
	\end{equation}
\end{proposition}

See Appendix \ref{appendix:A} for the detailed proof of the proposition. This proposition states that, unlike the commutative case, the Eq. \eqref{eq:2-12} has only zero solution in the context of the noncommutative case. For the real situation faced by the space-based gravitational wave detector like LISA,TianQin or Taiji, the proposition shows that there is no nonzero TDI combination that can completely suppress the laser frequency noise $p_i$. Therefore, it is necessary to explore alternative methods to mitigate the impact of laser frequency noise for the data and search for the so-called second-generation or higher-generation TDI combinations by other means.

\section{TDI of different generations}\label{sec:3}
In this section, we review the TDI of various generations and establish connections between TDI combinations of different generations.

In general, the armlengths $L_i(t)$ are functions of the time, TDI of different generations make different assumptions about $L_i(t)$. In the paper, we focus on TDI combinations of four different generations, i.e., the zeroth-generation TDI, the first-generation TDI, the modified first-generation TDI and the second-generation TDI.

\subsection{TDI with time-independent and equal armlengths}
For the most perfect case, three spacecrafts form a equilateral triangle, and the side length of the triangle remains the same. This is equivalent to the following assumption
\begin{equation}\label{eq:3-1-1}
	\begin{gathered}
		L_i(t) = L_i = L_j = L_j(t),\\
		i \in \{ 1, 2, 3 \},\; j \in \{ 1', 2', 3' \}.
	\end{gathered}
\end{equation}
Under the assumption \eqref{eq:3-1-1}, the six time-delay operators can all be written as the same one $D_1$. Eq. \eqref{eq:2-8}, \eqref{eq:2-11} and \eqref{eq:2-12} become
\begin{equation}\label{eq:3-1-2}
	M_0 =
	\begin{pmatrix}
		-1 & D_1 & 0 \\
		0 & -1 & D_1 \\
		D_1 & 0 & -1 \\
		-1 & 0 & D_1 \\
		D_1 & -1 & 0 \\
		0 & D_1 & -1
	\end{pmatrix},
\end{equation}

\begin{equation}\label{eq:3-1-3}
	\left\{
	\begin{aligned}
		q_1 + q_{1'} - D_1 q_{2'} - D_1 q_3 &= 0, \\
		q_2 + q_{2'} - D_1 q_{3'} - D_1 q_1 &= 0, \\
		q_3 + q_{3'} - D_1 q_{1'} - D_1 q_2 &= 0,
	\end{aligned}
	\right.
\end{equation}
and
\begin{equation}\label{eq:3-1-4}
	(1 - D_1^3) q_3 + D_1 (D_1 - 1) q_{1'} + D_1 (1 - D_1^2) q_{2'} + (1 - D_1^2) q_{3'} = 0.
\end{equation}

Solutions of Eq. \eqref{eq:3-1-3} or Eq. \eqref{eq:3-1-4} are called the zeroth-generation TDI solutions. According to the standard calculation process \cite{Ralf1996An,Cox2015Ideals}, we can get a generating set for the zeroth-generation TDI, such as
\begin{equation}\label{eq:3-1-5}
	\begin{aligned}
		g_1^{(0)} &= (1, 1, 1, -1, -1, -1), \\
		g_2^{(0)} &=(-D_1, 1, 0, 0, -1, D_1), \\
		g_3^{(0)} &=(-1, -1 - D_1, 0, 1 + D_1, 1, 0).
	\end{aligned}
\end{equation}

\subsection{TDI with time-independent and partially equal armlengths}
For the case that three spacecrafts form a scalene triangle, and the side lengths of the triangle remain the same, one has the following assumption
\begin{equation}\label{eq:3-2-1}
	\begin{gathered}
		L_1 \neq L_2 \neq L_3, \\
		L_i(t) = L_i = L_{i'}(t),\;	i \in \{ 1, 2, 3 \}.
	\end{gathered}
\end{equation}
Under the assumption \eqref{eq:3-2-1}, the time-delay operators are commutative and satisfy
\begin{equation}
	D_i = D_{i'}, \; i \in \{ 1, 2, 3 \}.
\end{equation}

Then Eq. \eqref{eq:2-8}, \eqref{eq:2-11} and \eqref{eq:2-12} become
\begin{equation}\label{eq:3-2-2}
	M_1 =
	\begin{pmatrix}
		-1 & D_3 & 0 \\
		0 & -1 & D_1 \\
		D_2 & 0 & -1 \\
		-1 & 0 & D_2 \\
		D_3 & -1 & 0 \\
		0 & D_1 & -1
	\end{pmatrix},
\end{equation}

\begin{equation}\label{eq:3-2-3}
	\left\{
	\begin{aligned}
		q_1 + q_{1'} - D_3 q_{2'} - D_2 q_3 &= 0, \\
		q_2 + q_{2'} - D_1 q_{3'} - D_3 q_1 &= 0, \\
		q_3 + q_{3'} - D_2 q_{1'} - D_1 q_2 &= 0,
	\end{aligned}
	\right.
\end{equation}
and
\begin{equation}\label{eq:3-2-4}
  (1 - D_1 D_2 D_3) q_3 + (D_1 D_3 - D_2) q_{1'} + D_1 (1 - D_3^2) q_{2'} + (1 - D_1^2) q_{3'} = 0.
\end{equation}

Solutions of Eq. \eqref{eq:3-2-3} or Eq. \eqref{eq:3-2-4} are called the first-generation TDI solutions. According to the standard calculation process \cite{Ralf1996An,Cox2015Ideals,Dhurandhar2002}, we can get a generating set for the first-generation TDI, such as the most common first-generation TDI combinations~\cite{Tinto2021Time}
\begin{equation}\label{eq:3-2-5}
	\begin{aligned}
		\alpha &= (-1, -D_3, -D_1 D_3, 1, D_1 D_2, D_2), \\
		\beta &= (-D_1 D_2, -1, -D_1, D_3, 1, D_2 D_3), \\
		\gamma &= (-D_2, -D_2 D_3, -1, D_1 D_3, D_1, 1), \\
		\zeta &= (-D_1, -D_2, -D_3, D_1, D_2, D_3).
	\end{aligned}
\end{equation}

\subsection{TDI with time-independent and unequal armlengths}
For the case that three spacecrafts do not form a equilateral triangle, if the Sagnac effect in rigid rotation is considered and the armlengths do not change with time, one have the following assumption
\begin{equation}\label{eq:3-3-1}
	\begin{gathered}
		L_1 \neq L_2 \neq L_3, \\
		L_i(t) = L_i \neq L_{i'} = L_{i'}(t),\; i \in \{ 1, 2, 3 \}.
	\end{gathered}
\end{equation}
Under the assumption \eqref{eq:3-3-1}, the six time-delay operators are different and commutative.

Then Eq. \eqref{eq:2-8} remains the same, and Eq. \eqref{eq:2-11} and \eqref{eq:2-12} become
\begin{equation}\label{eq:3-3-2}
	\left\{
	\begin{aligned}
		q_1 + q_{1'} - D_{3'} q_{2'} - D_2 q_3 &= 0, \\
		q_2 + q_{2'} - D_{1'} q_{3'} - D_3 q_1 &= 0, \\
		q_3 + q_{3'} - D_{2'} q_{1'} - D_1 q_2 &= 0,
	\end{aligned}
	\right.
\end{equation}
and
\begin{widetext}
  \begin{equation}\label{eq:3-3-3}
    (1 - D_1 D_2 D_3) q_3 + (D_1 D_3 - D_{2'}) q_{1'} + D_1 (1 - D_3 D_{3'}) q_{2'} + (1 - D_1 D_{1'}) q_{3'} = 0.
  \end{equation}
\end{widetext}

Solutions of Eq. \eqref{eq:3-3-2} or Eq. \eqref{eq:3-3-3} are called the modified first-generation or $1.5$-generation TDI solutions. According to the standard calculation process \cite{Ralf1996An,Cox2015Ideals,Rajesh2004}, we can also get a generating set for the modified first-generation TDI, such as
\begin{widetext}
  \begin{equation}\label{eq:3-3-4}
    \begin{aligned}
      g_1^{(1.5)} &= ( D_1 D_{1'} D_{3'} - D_{3'}, 1- D_3 D_{3'}, 0, 0, D_1 D_{1'} - 1, D_1 (1 - D_3 D_{3'}) ), \\
      g_2^{(1.5)} &= ( 1- D_1 D_{1'}, D_3 - D_{1'} D_{2'}, 0, D_1 D_{1'} - 1, 0, D_1 D_3 - D_{2'} ), \\
      g_3^{(1.5)} &= ( D_{1'} (1 - D_2 D_{2'}), 0, D_3 - D_{1'} D_{2'}, D_2 D_3 - D_{1'}, 0, D_3(D_2 D_{2'} - 1) ), \\
      g_4^{(1.5)} &= ( 0, 0, D_{1'} D_{2'} D_{3'} - 1, D_{1'} D_{3'} - D_2, D_{1'}(1 - D_2 D_{2'}), 1 - D_2 D_{2'} ), \\
      g_5^{(1.5)} &= ( 1- D_{1'} D_{2'} D_{3'}, 0, 0, D_3 D_{3'} - 1, D_3 - D_{1'} D_{2'}, D_{2'}(D_3 D_{3'} - 1) ), \\
      g_6^{(1.5)} &= ( 0, 1 - D_2 D_{2'}, D_1 - D_{2'} D_{3'}, D_1 D_2 - D_{3'}, D_2 D_{2'} - 1, 0 ).
    \end{aligned}
  \end{equation}
\end{widetext}

\subsection{TDI with time-dependent armlengths}
In a space-based gravitational wave detector, the armlengths between spacecrafts vary over time due to the inability to maintain constant distances. Generally, these armlengths $L_i(t)$ are functions that depend on time and can be expressed as the series expansions
\begin{equation}\label{eq:3-4-1}
	L_i(t) = L_i + t \dot{L}_i +\frac{t^2}{2} \ddot{L}_i + \cdots,
\end{equation}
where $\dot{L}_i$ and $\ddot{L}_i$ represent the first and second derivatives with respect to the time respectively. For the LISA mission, $\dot{L}_i$ is at most $10$ m/s~\cite{Pau2017Laser}. It has been demonstrated that the first-generation or modified first-generation TDI combinations do not meet the requirements of the noise suppression. Specifically, the estimated magnitude of the remaining laser frequency fluctuations in the center of the frequency band can be about 30 times higher than the level set by the secondary noise sources~\cite{Tinto2021Time}.

In cases of TDI with time-dependent armlengths, the six time-delay operators $D_i, D_{i'}$ are different and non-commutative, making the order in which they are applied important. According to Proposition \ref{prop:1}, no TDI combination can completely suppress the laser frequency noise $p_i$. In order to suppress the laser frequency noise below a level determined by secondary noises, we can only search for new TDI combinations that approximately satisfy Eq. \eqref{eq:2-9}.

In broad terms, a TDI combination is referred to as second-generation TDI if its effect on suppressing laser frequency noises is several orders of magnitude higher than that of the commonly used first-generation TDI combinations such as \eqref{eq:3-2-5}. For a second-generation TDI combination, substituting Eq. \eqref{eq:3-4-1} into Eq. \eqref{eq:2-5} and expanding the expression as a time-dependent series, the terms of $p_i$ in the series are zero or nearly zero.

Unlike the first-generation or modified first-generation TDI, there is no explicit equation, such as Eq.\eqref{eq:3-2-3} or \eqref{eq:3-3-2}, to represent the equations satisfied by the second-generation TDI due to Proposition \ref{prop:1}. Although an exhaustive search can be employed to find the second-generation TDI combination, the search space is considerably large due to the non-commutativity of the time-delay operator $D_i$. It is challenging to find new TDI combinations within an acceptable time frame. Therefore, to maximize the utilization of existing information in searching for new TDI combinations, we will establish connections between different generation TDI combinations in the following subsection and propose a search strategy for finding the higher-generation TDI using the generators of the lower-generation TDI in section \ref{sec:4}.

\subsection{Connections between TDI combinations of different generations}
First, we establish the connections between the second-generation TDI and the modified first-generation TDI. Since both noncommutative time-delay operators and commutative time-delay operators are involved here, it is necessary to define a mapping from noncommutative polynomials to commutative polynomials.

Let $ \boldsymbol{X} = \{ X_1, X_2, \dots, X_n \} $ and $ \boldsymbol{x} = \{ x_1, x_2, \dots, x_n \} $ be two alphabets. Let
\begin{equation}\label{eq:3-5-1}
	K \langle \boldsymbol{X} \rangle = K \langle X_1, X_2, \dots, X_n \rangle
\end{equation}
and
\begin{equation}\label{eq:3-5-2}
	K [\boldsymbol{x}] = K [ x_1, x_2, \dots, x_n ]
\end{equation}
denote the noncommutative polynomial ring and the commutative polynomial ring in $n$ variables with coefficients in the field $K$ respectively. We define a homomorphism $ \mathcal{F}_c : K \langle \boldsymbol{X} \rangle \to K [\boldsymbol{x}] $ such that it replaces $X_i$ as $x_i$. For example,
\begin{equation}\label{eq:3-5-3}
	\begin{aligned}
		\mathcal{F}_c (X_3 X_2 X_1 X_2 X_3) &= x_1 x_2^2 x_3^2, \\
		\mathcal{F}_c (X_2 X_1 X_3 + X_2 X_3 - X_3 X_1) &= x_1 x_2 x_3 + x_2 x_3 -x_1 x_3.
	\end{aligned}
\end{equation}
The subscript ``c'' is the abbreviation of ``commutative'', which means that the function $\mathcal{F}_c$ converts a noncommutative polynomial into a commutative polynomial.

For any $ a \in K[\boldsymbol{x}] $, the preimage $ \mathcal{F}_c^{-1} (a) $ of $a$ is defined by
\begin{equation}\label{eq:3-5-4}
	\mathcal{F}_c^{-1} (a) = \{ q \in K \langle \boldsymbol{X} \rangle \vert \mathcal{F}_c (q) = a \}.
\end{equation}
For any subset $ A \subset K[\boldsymbol{x}] $, the preimage $ \mathcal{F}_c^{-1} (A) $ of $A$ is defined by
\begin{equation}\label{eq:3-5-5}
	\mathcal{F}_c^{-1} (A) = \{ q \in K \langle \boldsymbol{X} \rangle \vert \mathcal{F}_c (q) \in A \}.
\end{equation}

Let
\begin{equation}\label{eq:3-5-6}
	K^m \langle \boldsymbol{X} \rangle = (a_1, \ldots, a_m),\; a_i \in K \langle \boldsymbol{X} \rangle,
\end{equation}
and
\begin{equation}\label{eq:3-5-7}
	K^m [\boldsymbol{x}] = (a_1, \ldots, a_m),\; a_i \in K [\boldsymbol{x}].
\end{equation}
The homomorphism $\mathcal{F}_c$ can be extented to the homomorphism $\bar{\mathcal{F}}_c$ that maps $K^m \langle \boldsymbol{X} \rangle$ to $K^m [\boldsymbol{x}]$, which is defined by
\begin{equation}\label{eq:3-5-8}
	\bar{\mathcal{F}}_c ( (a_1, \ldots, a_m) ) = ( \mathcal{F}_c(a_1), \ldots, \mathcal{F}_c(a_m) ).
\end{equation}

Similarly, the preimage of $ (a_1, \ldots, a_m) \in K^m[x] $ is defined by
\begin{equation}\label{eq:3-5-9}
	\bar{\mathcal{F}}_c^{-1} ( (a_1, \ldots, a_m) ) = \{ q \in K^6 \langle \boldsymbol{X} \rangle \vert \bar{\mathcal{F}}_c (q) = (a_1, \ldots, a_m) \}.
\end{equation}
For any subset $ A \subset K^m[x] $, the preimage $ \mathcal{F}_c^{-1} (A) $ of $A$ is defined by
\begin{equation}\label{eq:3-5-10}
	\bar{\mathcal{F}}_c^{-1} (A) = \{ q \in K^6 \langle \boldsymbol{X} \rangle \vert \mathcal{F}_c (q) \in A \}.
\end{equation}

In the context of the second-generation TDI, one has $m = 6$ and
\begin{equation}\label{eq:3-5-11}
	\begin{aligned}
		X_1 &= D_1,\; X_2 = D_2,\; X_3 = D_3, \\
		X_4 &= D_{1'},\; X_5 = D_{2'},\; X_6 = D_{3'}.
	\end{aligned}
\end{equation}
To simplify the statement, let
\begin{equation}\label{eq:3-5-12}
	\begin{aligned}
		\boldsymbol{D} &= \{ D_1, D_2, D_3, D_{1'}, D_{2'}, D_{3'} \}, \\
		\boldsymbol{x}_{6v} &= \{ x_1, x_2, x_3, x_{1'}, x_{2'}, x_{3'} \},
	\end{aligned}
\end{equation}
where the time-delay operators $D_i$ are denoted by $x_i$ for the modified first-generation TDI, and the subscript ``$6v$'' of $\boldsymbol{x}$ means that it contains six variables.

On the one hand, if $\boldsymbol{a} \in K^6 \langle \boldsymbol{D} \rangle$ is a second-generation TDI solution, then $ \bar{\mathcal{F}}_c (\boldsymbol{a}) \in K^6 [\boldsymbol{x}_{6v}]$ is a modified first-generation TDI solution, which satisfies Eq. \eqref{eq:3-3-2}. On the other hand, if $\boldsymbol{a} \in K^6 [\boldsymbol{x}_{6v}]$ is a modified first-generation TDI solution, then a combination in $\bar{\mathcal{F}}_c^{-1} (\boldsymbol{a})$ may be a second-generation TDI solution. Therefore, we can search the possible second-generation TDI combinations from the modified first-generation TDI combinations, and a search strategy for finding second-generation TDI combinations will be proposed in section \ref{sec:4}.

For completeness of the results, we also establish the connections between the modified first-generation TDI, the first generation TDI, and the zeroth-generation TDI. To avoid confusion, we use the notations $x_i$ to represent the time-delay operators $D_i$ when they are commutative in the paper.

Let
\begin{equation}\label{eq:3-5-13}
	\boldsymbol{x}_{3v} = \{ x_1, x_2, x_3 \},\; \boldsymbol{x}_{1v} = \{ x_1 \},
\end{equation}
where the subscript ``$3v$'' and ``$1v$'' are the abbreviation of ``3 variables'' and ``1 variable'' respectively. We can define a homomorphism $ \mathcal{F}_{3v} : K [\boldsymbol{x}_{6v}] \to K [\boldsymbol{x}_{3v}] $ such that it replaces $x_{i'}$ as $x_i$. For example,
\begin{equation}\label{eq:3-5-14}
	\mathcal{F}_{3v} (x_{1'} x_2 + x_1 x_{2'}) = 2 x_1 x_2.
\end{equation}
Similar to the Eq. \eqref{eq:3-5-4}, for any $ a \in K[\boldsymbol{x}_{3v}] $, the preimage $ \mathcal{F}_{3v}^{-1} (a) $ of $a$ is defined by
\begin{equation}\label{eq:3-5-15}
	\mathcal{F}_{3v}^{-1} (a) = \{ q \in K [\boldsymbol{x}_{6v}] \; \vert \; \mathcal{F}_{3v} (q) = a \}.
\end{equation}
The homomorphism $\mathcal{F}_{3v}$ can be extented to the homomorphism $\bar{\mathcal{F}}_{3v}$ that maps $K^6 [\boldsymbol{x}_{6v}]$ to $K^6 [\boldsymbol{x}_{3v}]$, which is defined by
\begin{equation}\label{eq:3-5-16}
	\bar{\mathcal{F}}_{3v} ( (a_1, \ldots, a_6) ) = ( \mathcal{F}_{3v}(a_1), \ldots, \mathcal{F}_c(a_6) ).
\end{equation}

Similarly, the preimage of $ (a_1, \ldots, a_6) \in K^6[x] $ is defined by
\begin{equation}\label{eq:3-5-17}
	\bar{\mathcal{F}}_{3v}^{-1} ( (a_1, \ldots, a_m) ) = \{ q \in K^6 [\boldsymbol{x}_{6v}] \; \vert \; \bar{\mathcal{F}}_{3v} (q) = (a_1, \ldots, a_6) \}.
\end{equation}

Thus, by using the homomorphism $\mathcal{F}_{3v}$, we establish the connections between the modified first-generation TDI combinations and the first-generation TDI combinations.

Similarly, we can define the $ \mathcal{F}_{1v} : K [\boldsymbol{x}_{3v}] \to K [\boldsymbol{x}_{1v}] $ such that it replaces $x_2$ and $x_3$ as $x_1$, and extent it to the homomorphism $\bar{\mathcal{F}}_{1v}$ that maps $K^6 [\boldsymbol{x}_{3v}]$ to $K^6 [\boldsymbol{x}_{1v}]$, which establishes the connections between the first-generation TDI combinations and the zeroth-generation TDI combinations.

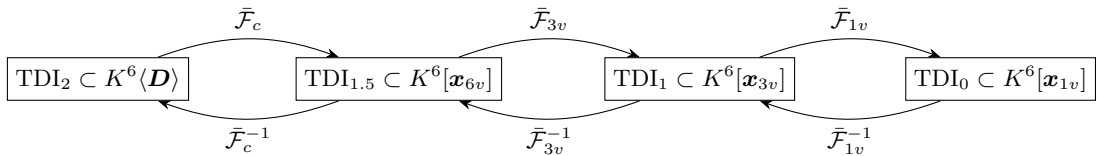
\begin{figure}[!ht]
	\centering
	\begin{tikzpicture}[>=Stealth, node distance=4cm]
		\node[rectangle, draw] (A) {$\mathrm{TDI}_2 \subset K^6 \langle \boldsymbol{D} \rangle$};
		\node[rectangle, draw, right of = A] (B) {$\mathrm{TDI}_{1.5} \subset K^6 [\boldsymbol{x}_{6v}]$};
		\node[rectangle, draw, right of = B] (C) {$\mathrm{TDI}_1 \subset K^6 [\boldsymbol{x}_{3v}]$};
		\node[rectangle, draw, right of = C] (D) {$\mathrm{TDI}_0 \subset K^6 [\boldsymbol{x}_{1v}]$};
		
		\draw[->] (A) to [out=20,in=160,looseness=1] node[above]{$\bar{\mathcal{F}}_c$} (B);
		\draw[->] (B) to [out=20,in=160,looseness=1] node[above]{$\bar{\mathcal{F}}_{3v}$} (C);
		\draw[->] (C) to [out=20,in=160,looseness=1] node[above]{$\bar{\mathcal{F}}_{1v}$} (D);
		\draw[->] (D) to [out=200,in=-20,looseness=1] node[below]{$\bar{\mathcal{F}}_{1v}^{-1}$} (C);
		\draw[->] (C) to [out=200,in=-20,looseness=1] node[below]{$\bar{\mathcal{F}}_{3v}^{-1}$} (B);
		\draw[->] (B) to [out=200,in=-20,looseness=1] node[below]{$\bar{\mathcal{F}}_c^{-1}$} (A);
		
	\end{tikzpicture}
	\caption{Connections between TDI combinations of different generations.}\label{fig:relation}
\end{figure}

The set of all TDI combinations of the $i$th-generation is denoted as $\mathrm{TDI}_i$, and the connections between TDI combinations of different generations are summarized in Fig. \ref{fig:relation}. We can transform higher-generation TDI combinations into lower-generation TDI combinations through the mappings $\bar{\mathcal{F}}_c$, $\bar{\mathcal{F}}_{3v}$, and $\bar{\mathcal{F}}_{1v}$, which also leave a path to find higher-generation TDI combinations through the preimage of lower-generation TDI combinations.

\section{A search strategy for higher-generation TDI combinations}\label{sec:4}
In this section, we propose a search strategy for identifying higher-generation TDI combinations, leveraging the connections established earlier between TDI combinations of different generations. Our approach is rooted in the principle that the image of every higher-generation TDI combination under a specific mapping corresponds to a relatively lower-generation TDI combination. Thus, we can explore higher-generation TDI combinations in the preimage of lower-generation ones.

Although we have already found the generators of the first-generation TDI and the modified first-generation TDI, we will demonstrate the effectiveness of the search strategy by searching for the first-generation TDI combinations using the generators of the zeroth-generation TDI, and searching for the modified first-generation TDI combinations using the generations of the first-generation TDI. The search strategy comprises four steps:
\begin{enumerate}
	\item Compute a generating set of solutions for the operator equation associated with the lower-generation TDI.
	\item Based on the generatiors in the generating set, construct a set $A_1$ of the lower-generation TDI combinations.
	\item Choose a finite set $A_2$ of combinations from the preimage $\bar{\mathcal{F}}^{-1}(A_1)$ of the set $A_1$ under a specific mapping.
	\item Search for higher-generation TDI combinations within the set $A_2$. If a suitable TDI combination is not found, iterate through the previous steps.
\end{enumerate}
We will explain this process step by step.

\subsection{Compute a generating set of solutions for the operator equation}
To obtain combinations for the zeroth-generation TDI, first-generation TDI, or modified first-generation TDI, we need to solve linear equations, such as those exemplified by Eq. \eqref{eq:3-1-4}, \eqref{eq:3-2-4}, or \eqref{eq:3-3-3}. The general process for obtaining the generators of solutions to linear equations over commutative polynomial rings is elaborated in Appendix \ref{appendix:B}. Different sets of the generators for the lower-generation TDI can be obtained depending on the choice of orderings, an aspect that has not been adequately addressed in previous literature. The impact of different generators on efficiency of the search strategy varies, emphasizing the necessity to consider generators acquired under different orderings. The orderings encountered during the process of solving the equations are introduced in Appendix \ref{appendix:C}, and the ordering ORD used in a specific solving process is recorded in the following form:
\begin{widetext}
  \begin{equation}\label{eq:4-1-1}
    \begin{aligned}
      \text{ORD} = \{\text{Variable ordering},\; \text{Monomial ordering},\; \text{Position ordering},\; \text{Relative ordering of terms and positions} \}.
    \end{aligned}
  \end{equation}
\end{widetext}

For example, the generating set \eqref{eq:3-1-5} for the zeroth-generation TDI is obtained under the ordering
\begin{equation}\label{eq:4-1-2}
	\text{ORD} = \{ x_1,\; \succ_\text{lex},\; \mathbf{e}_1 \succ \mathbf{e}_2 \succ \mathbf{e}_3 \succ \mathbf{e}_4,\; \text{TOP} \},
\end{equation}
and the generating set \eqref{eq:3-3-4} for the modified first-generation TDI is obtained under the ordering
\begin{equation}\label{eq:4-1-3}
	\text{ORD} = \{ x_1 \succ x_2 \succ x_3 \succ x_{1'} \succ x_{2'} \succ x_{3'},\; \succ_\text{lex},\; \mathbf{e}_1 \succ \mathbf{e}_2 \succ \mathbf{e}_3 \succ \mathbf{e}_4,\; \text{POT} \}.
\end{equation}

To streamline the computation of generators under different orderings, we implemented the procedure for solving linear equations over commutative polynomial rings, as outlined in Appendix \ref{appendix:B}. For the zeroth-generation TDI, the first-generation TDI, and the modified first-generation TDI, we calculated generating sets under 48, 864, and 103680 different orderings respectively (see Appendix \ref{appendix:C} for details). All calculations were executed on a CPU i5-12400 with six cores using parallel computing in Mathematica 13, and the results are summarized in Tables \ref{table:1}-\ref{table:3}.

\begin{table}[!ht]
  \centering
	\caption{Numbers of generating sets of the zeroth-generation TDI under 48 different orderings.}\label{table:1}
    \renewcommand\arraystretch{1.5}
    \begin{tabular}{ccccc}
      \toprule
      \toprule
      number of generators in the generating set & 3 \\
      \midrule
      number of orderings & 48 \\
      number of different generating sets & 13 \\
      \bottomrule
      \bottomrule
	\end{tabular}
\end{table}

\begin{table}[!ht]
    \centering
	\caption{Numbers of generating sets of generating sets of the first-generation TDI under 864 different orderings.}\label{table:2}
    \renewcommand\arraystretch{1.5}
    \begin{tabular}{ccccc}
      \toprule
      \toprule
      number of generators in the generating set & 4 & 5 & 6 & 4--6 \\
      \midrule
      number of orderings & 224 & 632 & 8 & 864 \\
      number of different generating sets & 42 & 18 & 8 & 68 \\
      \bottomrule
      \bottomrule
	\end{tabular}
\end{table}

\begin{widetext}
  \begin{table}[!ht]
    \centering
    \caption{Numbers of generating sets of the modified first-generation TDI under 103680 different orderings.}\label{table:3}
      \renewcommand\arraystretch{1.5}
      \begin{tabular}{ccccccc}
        \toprule
        \toprule
        number of generators in the generating set & 6 & 7 & 8 & 9 & 10 & 6--10 \\
        \midrule
        number of orderings & 16776 & 3384 & 7296 & 12696 & 63528 & 103680 \\
        number of different generating sets & 86 & 13 & 27 & 10 & 12 & 148 \\
        \bottomrule
        \bottomrule
    \end{tabular}
  \end{table}
\end{widetext}

For the zeroth-generation TDI, we calculate the generating sets of Eq. \eqref{eq:3-1-4} under 48 different orderings, taking about two seconds. It is found that each set contains 3 elements and there are 13 distinct generating sets in total, such as \eqref{eq:3-1-5}.

For the first-generation TDI, we calculate the generating sets of Eq. \eqref{eq:3-2-4} under 864 different orderings, taking about fifty seconds. It is found that these sets can contain between 4-6 generators, with 5 generators being the most common. A total of 68 distinct generating sets are identified. Interestingly, a previous study~\cite{Dhurandhar2002} used Macaulay2~\cite{Macaulay2} to obtain a generating set with 5 generators for the first-generation TDI, which were then manually reduced to 4 generators \eqref{eq:3-2-5}: $\alpha$, $\beta$, $\gamma$, and $\zeta$. While this set does not appear among the 68 sets we identified, each of the four generators is found within our generated sets. We can obtain a generating set with 4 generators under certain orderings. For example, under the ordering
\begin{equation}\label{eq:4-1-4}
	\text{ORD} = \{ x_1 \succ x_2 \succ x_3,\; \succ_\text{lex},\; \mathbf{e}_1 \succ \mathbf{e}_2 \succ \mathbf{e}_3 \succ \mathbf{e}_4,\; \text{POT} \},
\end{equation}
the generating set for the first-generation TDI contains
\begin{equation}\label{eq:4-1-5}
	\begin{gathered}
		(D_2, D_2 D_3, 1, -D_1 D_2, -D_1, -1), \\
		(1 - D_1 D_2 D_3, 0, 0, D_3^2 - 1, D_3 - D_1 D_2, D_2 D_3^2 - D_2), \\
		(D_1^2 D_3 - D_3, 1 - D_3^2, 0, 0, D_1^2 - 1, D_1 - D_1 D_3^2), \\
		(1 - D_1^2, D_3 - D_1 D_2, 0, D_1^2 - 1, 0, D_1 D_3 - D_2).
	\end{gathered}
\end{equation}

For the modified first-generation TDI, we calculate the generating sets of Eq. \eqref{eq:3-3-3} under an expansive 103680 orderings, taking approximately six hours. These sets potentially contain between 6-10 generators, with 10 being the most frequent. Despite the vast number of orderings considered, many yield the same results, resulting in 148 distinct generating sets. In literature \cite{Rajesh2004}, the authors obtained a generating set with 10 generators using CoCoA for the modified first-generation TDI, and then manually reduced to a generating set with 6 generators. It is able to obtain a set of 6 generators under certain orderings, such as ordering \eqref{eq:4-1-3} and generating set \eqref{eq:3-3-4}

Our findings allow us to reasonably determine the minimum number of elements a generating set must have for a specific generation TDI. The zeroth-generation, first-generation, and modified first-generation TDIs necessitate at least 3, 4, and 6 generators in a generating set, respectively.

\subsection{Construct a set of the lower-generation TDI combinations based on the generators}

Suppose that $G = \{ g_1, g_2, \ldots, g_n \}$ is a generating set of a specific generation TDI, the weighted summation of the generators
\begin{equation}\label{eq:4-2-1}
	\sum_{i=1}^n b_i g_i,\; b_i \in K[\boldsymbol{x}]
\end{equation}
is also an adequate TDI combination. Since there are infinitely many TDI combinations, we need to construct a set $A_1$ with finite elements using the generating set $G$. Here, we select a subset $B_1$ of the commutative polynomial ring $K[\boldsymbol{x}]$, and take the set $A_1$ as a subset of the set
\begin{equation}\label{eq:4-2-2}
	\left\{ \sum_{i=1}^n b_i g_i \; | \; b_i \in B_1 \subset K[\boldsymbol{x}] \right\}.
\end{equation}
The number of elements in set \eqref{eq:4-2-2} is approximately proportional to the $n$-th power of the number of elements in set $B_1$, and it is needed to select a subset from it to reduce the number of candidate combinations. For example, we can choose
\begin{equation}\label{eq:4-2-3}
	A_1 = \left\{ \sum_{i=1}^n b_i g_i \; | \; b_i \in B_1 \subset K[\boldsymbol{x}],\; \text{at most three of } b_i \text{ are nonzero} \right\}.
\end{equation}
The selection of sets $A_1$ and $B_1$ will affect the efficiency and results of the search strategy, as demonstrated below through examples.

For the zeroth-generation TDI, If we select \eqref{eq:3-1-5} as the generating set $G$ and
\begin{equation}\label{eq:4-2-4}
	B_1 = \{ 0, 1, -1, x_1, -x_1, 1 + x_1, -1 - x_1, 1 - x_1, x_1 - 1 \},
\end{equation}
one can expect to find the generators $\beta$, $\gamma$, and $\zeta$ of the first-generation TDI through the search strategy because of
\begin{equation}\label{eq:4-2-5}
	\begin{aligned}
		\bar{\mathcal{F}}_{1v}(\beta) &= x_1 g_1^{(0)} + (x_1 - 1) g_2^{(0)}, \\
		\bar{\mathcal{F}}_{1v}(\gamma) &= g_1^{(0)} + (x_1 - 1) g_3^{(0)}, \\
		\bar{\mathcal{F}}_{1v}(\zeta) &= x_1 g_1^{(0)}.
	\end{aligned}
\end{equation}
However, we can not find the generator $\alpha$ of the first-generation TDI because of
\begin{equation}\label{eq:4-2-6}
	\bar{\mathcal{F}}_{1v}(\alpha) = x_1^2 g_1^{(0)} + (x_1 - 1) g_2^{(0)} + (x_1 - 1) g_3^{(0)}
\end{equation}
and $x_1^2 \notin B_1$. In other words, we need to change either $G$ or $B_1$ in order to find $\alpha$. Specifically, if we select another generating set
\begin{equation}\label{eq:4-2-7}
	\begin{aligned}
		\alpha_0 &= (1, 0, -x_1, -1, x_1, 0) = - x_1 g_1^{(0)} - g_2^{(0)} - g_3^{(0)}, \\
		\beta_0  &= (-x_1, 1, 0, 0, -1, x) = g_2^{(0)}, \\
		\gamma_0 &= (0, -x_1, 1, x_1, 0, -1) = g_2^{(0)} + g_3^{(0)}, \\
		\zeta_0  &= (1, 1, 1, -1, -1, -1) = g_1^{(0)},
	\end{aligned}
\end{equation}
for the zeroth-generation TDI, but keep $B_1$ unchanged, one can find the complete generating set $\{ \alpha, \beta, \gamma, \zeta \}$ of the first-generation TDI through the search strategy because of
\begin{equation}\label{eq:4-2-8}
	\begin{aligned}
		\bar{\mathcal{F}}_{1v}(\alpha) &= \zeta_0 - x_1 \alpha_0 - \beta_0 - \gamma_0 = x_1 \zeta_0 + (1 - x_1)\alpha_0, \\
		\bar{\mathcal{F}}_{1v}(\beta)  &= \zeta_0 - \alpha_0 - x_1 \beta_0 - \gamma_0 = x_1 \zeta_0 + (1 - x_1)\beta_0, \\
		\bar{\mathcal{F}}_{1v}(\gamma) &= \zeta_0 - \alpha_0 - \beta_0 - x_1 \gamma_0 = x_1 \zeta_0 + (1 - x_1)\gamma_0, \\
		\bar{\mathcal{F}}_{1v}(\zeta)  &= \zeta_0 - \alpha_0 - \beta_0 - \gamma_0 = x\zeta_0.
	\end{aligned}
\end{equation}

This example illustrates that a generating set with fewer elements is not necessarily better to be used, and a generating set with more elements may have some advantages for the search strategy.

\subsection{Choose a finite set of combinations from the preimage}
It should be noted that the functions $\mathcal{F}_c$, $\mathcal{F}_{3v}$, and $\mathcal{F}_{1v}$ defined in section \ref{sec:3} are not injective, and $\mathcal{F}_c^{-1}(0)$, $\mathcal{F}_{3v}^{-1}(0)$, and $\mathcal{F}_{1v}^{-1}(0)$ all contain an infinite number of elements. The functions $\bar{\mathcal{F}}_c$, $\bar{\mathcal{F}}_{3v}$, and $\bar{\mathcal{F}}_{1v}$ have the same properties. Hence we need to carefully select a finite number of combinations in the preimage of a TDI combination under the mappings to make the search strategy feasible.

For a polynomial $a$, we select from the set $\mathcal{F}^{-1}(a)$ polynomials with the same number and form of monomials as $a$, and denote such a set as $\mathcal{H}^{-1}(a)$. For example,
\begin{equation}\label{eq:4-3-1}
	\begin{aligned}
		\mathcal{H}_{1v}^{-1}(x_1^2) &= \{ x_1 x_2, x_1 x_3, x_2 x_3 \} \subset \mathcal{F}_{1v}^{-1}(x_1^2), \\
		\mathcal{H}_{3v}^{-1}(x_1 x_2) &= \{ x_1 x_2, x_1 x_{2'}, x_{1'} x_2, x_{1'} x_{2'} \} \subset \mathcal{F}_{3v}^{-1}(x_1 x_2), \\
		\mathcal{H}_{c}^{-1}(x_1 x_2) &= \{ D_1 D_2 , D_2 D_1 \} \subset \mathcal{F}_{c}^{-1}(x_1 x_2), \\
	\end{aligned}
\end{equation}
and 
\begin{widetext}
  \begin{equation}\label{eq:4-3-2}
    \mathcal{H}_{c}^{-1}(x_1 x_2 + x_2 x_3) = \{ D_1 D_2 + D_2 D_3, D_1 D_2 + D_3 D_2, D_2 D_1 + D_2 D_3, D_2 D_1 + D_3 D_2 \}.
  \end{equation}
\end{widetext}

For a lower-generation TDI combination $\boldsymbol{a}$, with the aid of $\mathcal{H}^{-1}(a)$, we can select a set $\bar{\mathcal{H}}^{-1}(\boldsymbol{a})$ consisting of finite terms from $\bar{\mathcal{F}}^{-1}(\boldsymbol{a})$. Specifically, suppose that 
\begin{equation}\label{eq:4-3-3}
	\boldsymbol{q} = ( a_1, a_2, a_3, a_4, a_5, a_6 )
\end{equation}
is a TDI combination and $a_i$ are polynomials. We define
\begin{widetext}
  \begin{equation}\label{eq:4-3-4}
    \begin{aligned}
      \bar{\mathcal{H}}^{-1}(\boldsymbol{q}) =& \{ ( q_1, q_2, q_3, q_{1'}, q_{2'}, q_{3'} ) \;|\; q_3 \in \mathcal{H}^{-1}(a_3), q_{1'} \in \mathcal{H}^{-1}(a_4), q_{2'} \in \mathcal{H}^{-1}(a_5),\\
      &\;\; q_{3'} \in \mathcal{H}^{-1}(a_6), q_1\; \text{and}\; q_2\; \text{are calculated by using Eq. \eqref{eq:3-1-3}, \eqref{eq:3-2-3} or \eqref{eq:3-3-2}}  \},
    \end{aligned}
  \end{equation}
\end{widetext}
where
\begin{equation}\label{eq:4-3-5}
	\begin{aligned}
		q_1 &= -q_{1'} + x_{3} q_{2'} + x_2 q_3,\\
		q_2 &= -q_{2'} + x_{1} q_{3'} + x_3 q_1
	\end{aligned}
\end{equation}
for the first-generation TDI,
\begin{equation}\label{eq:4-3-6}
	\begin{aligned}
		q_1 &= -q_{1'} + x_{3'} q_{2'} + x_2 q_3,\\
		q_2 &= -q_{2'} + x_{1'} q_{3'} + x_3 q_1
	\end{aligned}
\end{equation}
for the modified first-generation TDI, and
\begin{equation}\label{eq:4-3-7}
	\begin{aligned}
		q_1 &= -q_{1'} + q_{2'} D_{3'} + q_3 D_2,\\
		q_2 &= -q_{2'} + q_{3'} D_{1'} + q_1 D_3
	\end{aligned}
\end{equation}
for the second-generation TDI.

Thus, we obtain a set with finite candidates
\begin{equation}\label{eq:4-3-8}
	A_2 = \bar{\mathcal{H}}^{-1}(A_1) = \{ \boldsymbol{q} \in \bar{\mathcal{H}}^{-1}(\boldsymbol{a}) \;|\; \boldsymbol{a} \in A_1 \}.
\end{equation}

\subsection{Search for higher-generation TDI combinations}
By following the preceding three steps, we obtain the set $A_2$ consisting of finite elements. The goal is to identify high-generation TDI combinations within $A_2$. The first or modified first-generation TDI can be easily validated against Eq. \eqref{eq:3-1-4} or \eqref{eq:3-2-4}. However, to confirm whether a combination qualifies as a second-generation TDI combination, the laser frequency noise $p_i (i=1,2,3)$ present in the TDI combination usually needs to be expanded to the first-order term with respect to time \cite{Dhurandhar2008General,Tinto2021Time,Wu2023Combinatorial}. If $A_2$ does not contain a suitable TDI combination, the aforementioned three steps can be repeated to generate a new set, after which the evaluation process can be resumed. It is important to note that the size of set $A_2$ can become exceedingly large, necessitating a judicious choice of the generating set and $A_1$ to limit the number of potential combinations.

It should be noted that if one wishes to employ this search strategy to develop a practical algorithm for searching for second-generation TDI combinations, each of these four steps must be optimized. Otherwise, there will be an excessive number of candidate combinations to evaluate. Below is an example of a second-generation TDI combination found using this search strategy. If \eqref{eq:3-3-4} is chosen as the generating set of the modified first-generation TDI and
\begin{equation}\label{eq:4-4-1}
	B_1 = \{ 0, 1 - x_3 x_{1'} x_{2'} x_{3'}^2 \}.
\end{equation}
By following the search strategy, we can identify at least one second-generation TDI combination
\begin{equation}\label{eq:4-4-2}
	\boldsymbol{q}_1 = ( q_1, q_2, q_3, q_{1'}, q_{2'}, q_{3'} ),
\end{equation}
where
\begin{equation}\label{eq:4-4-3}
	\begin{aligned}
		q_1 &= 1 - D_{2'} D_{1'} D_{3'} - D_{2'} D_{1'} D_{3'} D_{3} D_{3'}\\
		&\quad + D_{3} D_{3'} D_{2'} D_{1'} D_{3'} D_{2'} D_{1'} D_{3'},\\
		q_2 &= q_3 = 0,\\
		q_{1'} &= -1 + D_{3} D_{3'} + D_{3} D_{3'} D_{2'} D_{1'} D_{3'}\\
		&\quad - D_{2'} D_{1'} D_{3'} D_{3} D_{3'} D_{3} D_{3'},\\
		q_{2'} &= D_3 - D_{2'} D_{1'} + D_{3'} D_{3} D_{2'} D_{1'} - D_{2'} D_{1'} D_{3'} D_{3} \\
		&\quad - D_{2'} D_{1'} D_{3'} D_{3} D_{3'} D_{3} + D_{3'} D_{3} D_{2'} D_{1'} D_{3'} D_{2'} D_{1'}, \\
		q_{3'} &= - D_{2'} + D_{3} D_{3'} D_{2'} + D_{3} D_{3'} D_{2'} D_{1'} D_{3'} D_{2'}\\
		&\quad - D_{2'} D_{1'} D_{3'} D_{3} D_{3'} D_{3} D_{3'} D_{2'},
	\end{aligned}
\end{equation}
and
\begin{equation}\label{eq:4-4-4}
	\bar{\mathcal{F}}_c (\boldsymbol{q}_1) = (1 - x_3 x_{1'} x_{2'} x_{3'}^2) g_5^{(1.5)}.
\end{equation}

\begin{figure}[!ht]
	\centering
	\includegraphics[width=0.45\textwidth]{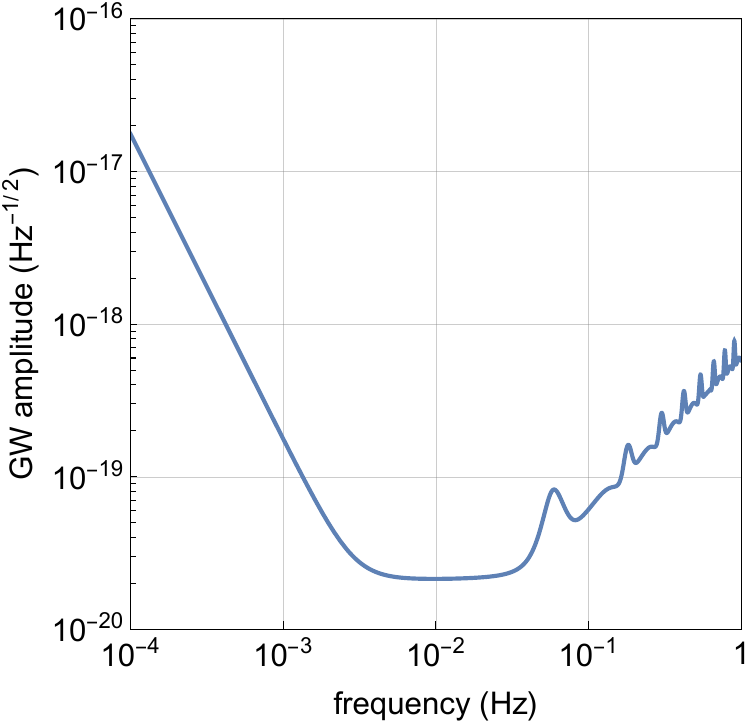}
	\caption{The sensitivity curve of the second-generation TDI combination $\boldsymbol{q}_1$.}\label{fig:sensitivity}
\end{figure}

This TDI combination seems not to have been reported in other literatures. To verify that $\boldsymbol{q}_1$ is indeed a second-generation TDI combination, one has
\begin{widetext}
  \begin{equation}\label{eq:4-4-5}
    \begin{aligned}
      \boldsymbol{q}_1 &= q_1 \eta_1 + q_{1'} \eta_{1'} + q_{2'} \eta_{2'} + q_{3'} \eta_{3'}\\
      &= (q_1 + q_{1'} - D_{3'} q_{2'})p_1 + (q_{2'} - D_{1'} q_{3'} - D_3 q_1)p_2 + (q_{3'} - D_{2'} q_{1'})p_3\\
      &= ( D_{2'} D_{1'} D_{3'} D_{3} D_{3'} D_{3} D_{3'} D_{2'} D_{1'} - D_{3} D_{3'} D_{2'} D_{1'} D_{3'} D_{2'} D_{1'} D_{3'} D_{3} )p_2 \\
      &\approx k_1 k_2 \dot{p}_2(t - 2L_3 - 2L_{1'} - 2L_{2'} - 3L_{3'}),
    \end{aligned}
  \end{equation}
\end{widetext}
where
\begin{equation}\label{eq:4-4-6}
	\begin{aligned}
		k_1 &= \dot{L}_{3} + \dot{L}_{1'} + \dot{L}_{2'} + 2\dot{L}_{3'} + \text{higher order terms},\\
		k_2 &= (L_{3} \dot{L}_{1'} + L_{3'} \dot{L}_{1'} + L_{3} \dot{L}_{2'} + L_{3} \dot{L}_{3'} + L_{3'} \dot{L}_{2'})\\
		&\quad -(L_{1'} \dot{L}_{3} + L_{1'} \dot{L}_{3'} + L_{2'} \dot{L}_{3} + L_{2'} \dot{L}_{3'} + L_{3'} \dot{L}_{3})\\
		&\quad + \text{higher order terms},
	\end{aligned}
\end{equation}
By using the LISA mission's typical parameters~\cite{Babak2021LISA}, the corresponding sensitivity curve \cite{Tinto2021Time,Wang2021Sensitivity,Wang2022Geometric} of $\boldsymbol{q}_1$ is showed in Fig. \ref{fig:sensitivity}.

\subsection{Differences and connections with the related approach}
In recent work \cite{Tinto2023Second}, the authors proposed a method to obtain the second-generation TDI from the first-generation TDI using  "lifting procedure". This method establishes a homomorphism between the first-generation TDI space and the second-generation TDI space. Here, we will restate this lifting procedure using the symbols provided in the paper.

The lifting procedure in \cite{Tinto2023Second} is divided into 3 steps:
\begin{enumerate}
	\item For a first-generation TDI combination $a$, select a special $\bar{a} \in \bar{\mathcal{F}}_{6v}^{-1}\bar{\mathcal{F}}_{3v}^{-1}(a)$ and decompose it into two Doppler measurements $a_{\uparrow}$ and $a_{\downarrow}$.
	\item ``Lift'' $a_{\uparrow}$ and $a_{\downarrow}$ to $a_{\uparrow\uparrow}$ and $a_{\downarrow\downarrow}$ by applying a iterative procedure.
	\item Combine $a_{\uparrow\uparrow}$ and $a_{\downarrow\downarrow}$ to obtain a second-generation TDI combination.
\end{enumerate}

In particular, the lifting procedure is elucidated through the illustration of the Sagnac combination $\alpha$ herein (see IV.B in \cite{Tinto2023Second}). To maintain consistency with the notation employed in \cite{Tinto2023Second}, the $\alpha$ utilized below differs from the $\alpha$ in the paper by a sign. The Sagnac combination $\alpha \in \mathrm{TDI}_1$ can be written as follows:
\begin{equation}\label{eq:4-5-1}
    \begin{aligned}
        \alpha &= (1, D_3, D_1 D_3, -1, -D_1 D_2, -D_2) \\
        &= (1, D_3, D_1 D_3, 0, 0, 0) - (0, 0, 0, 1, D_1 D_2, D_2).
    \end{aligned}
\end{equation}
According to the geometric considrations on the delays and paths traveled by the two synthesized beams, we can choose a special $\bar{\alpha} \in \bar{\mathcal{F}}_{6v}^{-1}\bar{\mathcal{F}}_{3v}^{-1}(\alpha)$ such that
\begin{equation}\label{eq:4-5-2}
    \bar{\alpha} = a_{\uparrow} - a_{\downarrow},
\end{equation}
where
\begin{equation}\label{eq:4-5-3}
    \begin{aligned}
        a_{\uparrow} &= (1, D_3, D_3 D_1, 0, 0, 0),\\
        a_{\downarrow} &= (0, 0, 0, 1, D_{2'} D_{1'}, D_{2'}).
    \end{aligned}
\end{equation}
The expression of the residual laser noise $p_1(t)$ in $\bar{\alpha}$ is
\begin{equation}\label{eq:4-5-4}
    \begin{aligned}
        (\bar{\alpha})_\text{\it res} &= (a_{\uparrow})_\text{\it res} - (a_{\downarrow})_\text{\it res} \\
        &= (D_3 D_1 D_2 - I)p_1 - (D_{2'} D_{1'} D_{3'} - I)p_1 \\
        &= (D_3 D_1 D_2 - D_{2'} D_{1'} D_{3'})p_1.
    \end{aligned}
\end{equation}

By applying a iterative procedure, one can define
\begin{equation}\label{eq:4-5-5}
    \begin{aligned}
        a_{\uparrow\uparrow} &= D_{2'} D_{1'} D_{3'} a_{\uparrow} + a_{\downarrow},\\
        a_{\downarrow\downarrow} &= a_{\uparrow} + D_3 D_1 D_2 a_{\downarrow},
    \end{aligned}
\end{equation}
which imply
\begin{equation}\label{eq:4-5-6}
    \begin{aligned}
        (a_{\uparrow\uparrow})_\text{\it res} &= (D_{2'} D_{1'} D_{3'} D_3 D_1 D_2 - I)p_1,\\
        (a_{\downarrow\downarrow})_\text{\it res} &= (D_3 D_1 D_2 D_{2'} D_{1'} D_{3'} - I)p_1.
    \end{aligned}
\end{equation}
Thus, we obtain a second-generation TDI combination
\begin{equation}\label{eq:4-5-7}
    \alpha_2 = (D_3 D_1 D_2 D_{2'} D_{1'} D_{3'} - I) a_{\uparrow\uparrow} - (D_{2'} D_{1'} D_{3'} D_3 D_1 D_2 - I) a_{\downarrow\downarrow} \in \mathrm{TDI}_2,
\end{equation}
with its residual laser noise being given by
\begin{equation}\label{eq:4-5-8}
    (\alpha_2)_\text{\it res} = [D_{2'} D_{1'} D_{3'} D_3 D_1 D_2, D_3 D_1 D_2 D_{2'} D_{1'} D_{3'}]p_1.
\end{equation}

Using the lifting procedure, the authors in \cite{Tinto2023Second} obtained the second-generation TDI combinations, denoted as $\alpha_2$, $\beta_2$, $\gamma_2$, and $X_2$, corresponding to the generators $\alpha$, $\beta$, $\gamma$ and the unequal-arm Michelson combination $X$ of the first-generation TDI. This establishes a homomorphism
\begin{equation}\label{eq:4-5-9}
    \psi: \mathrm{TDI}_1 \to \mathrm{TDI}_2,
\end{equation}
which maps each first-generation TDI combination $a$ to a second-generation TDI combinations
\begin{equation}\label{eq:4-5-10}
    a_2 = \psi(a) = \lambda_X X_2 + \lambda_\alpha \alpha_2 + \lambda_\beta \beta_2 + \lambda_\gamma \gamma_2.
\end{equation}

In summary, literature \cite{Tinto2023Second} employs an lifting procedure to generate second-generation TDI combinations from the first-generation counterparts. In contrast, this paper establishes a homomorphism from the second-generation TDI space to the modified first-generation TDI space, searching for second-generation TDI combinations within the preimages of the modified first-generation TDI combinations. While the approach presented in \cite{Tinto2023Second} yields an infinite variety of second-generation TDI combinations, it may not contain all possible second-generation TDI combinations through the lifting procedure. The search strategy proposed in this paper serves as a complementary approach, aiming to encompass a broader spectrum of second-generation TDI combinations.

\section{Equivalent TDI combinations}\label{sec:5}
According to the symmetry of the spacecrafts of the space-based gravitational wave detector, many TDI combinations exhibit similar performance for suppressing laser frequency noise \cite{Hartwig2022Characterization}. In order to simplify the number of TDI combinations and clarify the relationships between them, we explicitly define equivalent TDI combinations in the section.

Based on a specific TDI combination, one can construct up to 12 different equivalent TDI combinations based on the following three transformations:
\begin{enumerate}
	\item Rotate the indices of spacecrafts such as $1 \to 2 \to 3 \to 1$;
	\item Exchange two indices of spacecrafts such as $2 \leftrightarrow 3$;
	\item Add a minus sign to the whole TDI solution.
\end{enumerate}

These TDI combinations may have different forms and are considered equivalent in some sense. Specifically, we explicitly define the algebraic forms of the first two transformations and demonstrate the properties of the TDI solutions in the rest of the section.

For convenience, we redefine some notations appearing in section \ref{sec:2}. The 2-tuple $(i,j)$ consists of the indices of the spacecrafts, satisfying $i,j \in \{ 1, 2, 3 \}$ and $i \neq j$. The auxiliary variables $\eta_{(i,j)}$, the armlengths $L_{(i,j)}$ and the time-delay operators $D_{(i,j)}$ are denoted by
\begin{widetext}
  \begin{equation}\label{eq:5-1}
    \begin{aligned}
      \eta_{(2,1)} &= \eta_{1}, \eta_{(3,2)} = \eta_{2}, \eta_{(1,3)} = \eta_{3}, \eta_{(3,1)} = \eta_{1'}, \eta_{(1,2)} = \eta_{2'}, \eta_{(2,3)} = \eta_{3'}, \\
      L_{(3,2)} &= L_{1}, L_{(1,3)} = L_{2}, L_{(2,1)} = L_{3}, L_{(2,3)} = L_{1'}, L_{(3,1)} = L_{2'}, L_{(1,2)} = L_{3'}, \\
      D_{(3,2)} &= D_{1}, D_{(1,3)} = D_{2}, D_{(2,1)} = D_{3}, D_{(2,3)} = D_{1'}, D_{(3,1)} = D_{2'}, D_{(1,2)} = D_{3'}
    \end{aligned}
  \end{equation}
\end{widetext}
Thus, Eq. \eqref{eq:2-4} is simplified to
\begin{equation}\label{eq:5-2}
	\eta_{(i,j)} = D_{(i,j)} p_i - p_j.
\end{equation}

We define the mapping $\mathcal{F}_{RI}$ that it rotates the indices $(i,j)$ of spacecrafts and indices $i$ of the laser frequency noise $p_i$ in the expression as $1 \to 2 \to 3 \to 1$, which the subscript ``RI'' means ``Rotate Indices''. For example,
\begin{equation}\label{eq:5-3}
	\begin{aligned}
		\mathcal{F}_{RI}(\eta_{(i,j)}) &= D_{(i+1,j+1)} p_{i+1} - p_{j+1} = \eta_{(i+1,j+1)}, \\
		\mathcal{F}_{RI}(D_3 \eta_1) &= \mathcal{F}_{RI}(D_{(2,1)} \eta_{(2,1)}) = D_{(3,2)} \eta_{(3,2)} = D_1 \eta_2, \\
		\mathcal{F}_{RI}( \boldsymbol{\eta} ) &= (\eta_2, \eta_3, \eta_1, \eta_{2'}, \eta_{3'}, \eta_{1'})^T = M_{RI} \boldsymbol{\eta},
	\end{aligned}
\end{equation}
where
\begin{equation}\label{eq:5-4}
	M_{RI} =
	\begin{pmatrix}
		0 & 1 & 0 & 0 & 0 & 0 \\
		0 & 0 & 1 & 0 & 0 & 0 \\
		1 & 0 & 0 & 0 & 0 & 0 \\
		0 & 0 & 0 & 0 & 1 & 0 \\
		0 & 0 & 0 & 0 & 0 & 1 \\
		0 & 0 & 0 & 1 & 0 & 0
	  \end{pmatrix}
\end{equation}

Let
\begin{equation}\label{eq:5-5}
	\begin{aligned}
		q_1 &= q_1 (D_1, D_2 ,D_3 ,D_{1'}, D_{2'}, D_{3'}) \\
			&= q_1 (D_{(3,2)}, D_{(1,3)} ,D_{(2,1)} ,D_{(2,3)}, D_{(3,1)}, D_{(1,2)}).
	\end{aligned}
\end{equation}
Then, we have
\begin{equation}\label{eq:5-6}
	\begin{aligned}
		\mathcal{F}_{RI}(q_1) &= q_1 (D_{(1,3)}, D_{(2,1)} ,D_{(3,2)} ,D_{(3,1)}, D_{(1,2)}, D_{(2,3)}) \\
		&= q_1 (D_2, D_3 ,D_1 ,D_{2'}, D_{3'}, D_{1'}),
	\end{aligned}
\end{equation}
which implies
\begin{equation}\label{eq:5-7}
	\begin{aligned}
		\mathcal{F}_{RI}(\mathrm{TDI}) &= \mathcal{F}_{RI}(\mathbf{q}^T \boldsymbol{\eta}) = (\mathcal{F}_{RI}(q_1), \dots, \mathcal{F}_{RI}(q_{3'})) \mathcal{F}_{RI}(\boldsymbol{\eta}) \\
		&= (\mathcal{F}_{RI}(q_1), \dots, \mathcal{F}_{RI}(q_{3'})) M_{RI} M_2 \mathbf{p}
	\end{aligned}
\end{equation}
where
\begin{equation}\label{eq:5-8}
	M_{RI} M_2 =
	\begin{pmatrix}
		0 & -1 & D_1 \\
		D_2 & 0 & -1 \\
		-1 & D_3 & 0 \\
		D_{3'} & -1 & 0 \\
		0 & D_{1'} & -1 \\
		-1 & 0 & D_{2'}
	\end{pmatrix}.
\end{equation}

We can define the mapping $\mathcal{F}_{EI}$ that it exchange the indices $(i,j)$ of spacecrafts and indices $i$ of the laser frequency noise $p_i$ in the expression as $2 \leftrightarrow 3$, which the subscript ``EI'' means ``Exchange Indices''. For example,
\begin{equation}\label{eq:5-9}
	\begin{aligned}
		\mathcal{F}_{EI}(D_3 \eta_1) &= \mathcal{F}_{EI}(D_{(2,1)} \eta_{(2,1)}) = D_{(3,1)} \eta_{(3,1)} = D_{2'} \eta_{1'}, \\
		\mathcal{F}_{EI}( \boldsymbol{\eta} ) &= (\eta_{1'}, \eta_{3'}, \eta_{2'}, \eta_1, \eta_3, \eta_2)^T = M_{EI} \boldsymbol{\eta},
	\end{aligned}
\end{equation}
where
\begin{equation}\label{eq:5-10}
	M_{EI} =
	\begin{pmatrix}
		0 & 0 & 0 & 1 & 0 & 0 \\
		0 & 0 & 0 & 0 & 0 & 1 \\
		0 & 0 & 0 & 0 & 1 & 0 \\
		1 & 0 & 0 & 0 & 0 & 0 \\
		0 & 0 & 1 & 0 & 0 & 0 \\
		0 & 1 & 0 & 0 & 0 & 0
	  \end{pmatrix}.
\end{equation}

Thus, we have
\begin{equation}\label{eq:5-11}
	\mathcal{F}_{EI}(\mathrm{TDI}) = (\mathcal{F}_{EI}(q_1), \dots, \mathcal{F}_{EI}(q_{3'})) M_{EI} M_2 \mathbf{p}
\end{equation}
where
\begin{equation}\label{eq:5-12}
	M_{EI} M_2 =
	\begin{pmatrix}
		-1 & 0 & D_{2'} \\
		0 & D_{1'} & -1 \\
		D_{3'} & -1 & 0 \\
		-1 & D_3 & 0 \\
		D_2 & 0 & -1 \\
		0 & -1 & D_1
	\end{pmatrix}.
\end{equation}

Therefore, for any TDI combination $a$, based on the three transformations, one can construct up to 12 different equivalent TDI combinations:
\begin{widetext}
  \begin{equation}\label{eq:5-13}
    \begin{gathered}
      a,\; \mathcal{F}_{RI}(a),\; \mathcal{F}_{RI}^2(a),\; \mathcal{F}_{EI}(a),\; \mathcal{F}_{RI} \mathcal{F}_{EI}(a),\; \mathcal{F}_{RI}^2 \mathcal{F}_{EI}(a), \\
      -a,\; -\mathcal{F}_{RI}(a),\; -\mathcal{F}_{RI}^2(a),\;	-\mathcal{F}_{EI}(a),\; -\mathcal{F}_{RI} \mathcal{F}_{EI}(a),\; -\mathcal{F}_{RI}^2 \mathcal{F}_{EI}(a).
    \end{gathered}
  \end{equation}
\end{widetext}

For example, if
\begin{widetext}
  \begin{equation}\label{eq:5-14}
    \begin{aligned}
      a &= ( 1 - D_{2'}D_{1'}D_{3'}, (1 - D_{2'}D_{1'}D_{3'})D_3, (1 - D_{2'}D_{1'}D_{3'})D_3D_1, \\
      &\qquad D_3D_1D_2 - 1, (D_3D_1D_2 - 1)D_{2'}D_{1'}, (D_3D_1D_2 - 1)D_{2'} )
    \end{aligned}
  \end{equation}
\end{widetext}
is the second-generation Michelson combination~\cite{Wang2022Geometric}, one has
\begin{widetext}
  \begin{equation}\label{eq:5-15}
    \begin{aligned}		
      \mathcal{F}_{RI}(a) &= ( (1 - D_{3'}D_{2'}D_{1'})D_1D_2, 1 - D_{3'}D_{2'}D_{1'}, (1 - D_{3'}D_{2'}D_{1'})D_1, \\
      &\qquad (D_1D_2D_3 - 1)D_{3'}, D_1D_2D_3 - 1, (D_1D_2D_3 - 1)D_{3'}D_{2'} ),\\
      \mathcal{F}_{RI}^2(a) &= ( (1 - D_{1'}D_{3'}D_{2'})D_2, (1 - D_{1'}D_{3'}D_{2'})D_2D_3, 1 - D_{1'}D_{3'}D_{2'}, \\
      &\qquad (D_2D_3D_1 - 1)D_{1'}D_{3'}, (D_2D_3D_1 - 1)D_{1'}, D_2D_3D_1 - 1 ),\\
      \mathcal{F}_{EI}(a) &= ( D_2D_1D_3 - 1, (D_2D_1D_3 - 1)D_{3'}, (D_2D_1D_3 - 1)D_{3'}D_{1'}, \\
      &\qquad 1 - D_{3'}D_{1'}D_{2'}, (1 - D_{3'}D_{1'}D_{2'})D_2D_1, (1 - D_{3'}D_{1'}D_{2'})D_2 ),\\
      \mathcal{F}_{RI} \mathcal{F}_{EI}(a) &= ( (D_3D_2D_1 - 1)D_{1'}D_{2'}, D_3D_2D_1 - 1, (D_3D_2D_1 - 1)D_{1'}, \\
      &\qquad (1 - D_{1'}D_{2'}D_{3'})D_3, 1 - D_{1'}D_{2'}D_{3'}, (1 - D_{1'}D_{2'}D_{3'})D_3D_2 ),\\
      \mathcal{F}_{RI}^2 \mathcal{F}_{EI}(a) &= ( (D_1D_2D_3 - 1)D_{2'}, (D_1D_3D_2 - 1)D_{2'}D_{3'}, D_1D_3D_2 - 1, \\
      &\qquad (1 - D_{2'}D_{3'}D_{1'})D_1D_3, (1 - D_{2'}D_{3'}D_{1'})D_1, 1 - D_{2'}D_{3'}D_{1'} ).
    \end{aligned}
  \end{equation}
\end{widetext}

In the previous section, we obtain lots of generating sets of the zeroth-generation TDI, the first-generation TDI and the modified first-generation TDI under different orderings, but many of generations within these sets are equivalent. We will select non-equivalent TDI combinations within the different generating sets, for future use in the search strategy proposed in Section \ref{sec:4}.

For the zeroth-generation TDI, as listed in Table \ref{table:1}, we obtain 13 distinct generating sets under 48 different orderings, collectively comprising 39 generators. After removing equivalent TDI combinations, these generators are reduced to 5 inequivalent generators, as listed in Table \ref{table:4}.
\begin{table}[!ht]
  \centering
  \caption{Inequivalent generators of the zeroth-generation TDI in 13 different generating sets.}\label{table:4}
  \renewcommand\arraystretch{1.5}
  \begin{tabular}{c|c}
  \toprule
  \toprule
  No. & generator\\
  \midrule
    1 & $(-1, -1, -1, 1, 1, 1)$ \\
	  2 & $(-1, 0, D_1, 1, -D_1, 0)$ \\
	  3 & $(-1, -1 - D_1, 0, 1 + D_1, 1, 0)$ \\
		4 & $(-1 - D_1, -D_1, 0, 1 + D_1, 0, D_1)$ \\
		5 & $(-1 - D_1, -D_1 - D_1^2, D_1, 1 + D_1 + D_1^2, 0, 0)$ \\
  \bottomrule
  \bottomrule
  \end{tabular}
  \end{table}

For the first-generation TDI, as listed in Table \ref{table:2}, we obtain 68 distinct generating sets under 864 different orderings, collectively comprising 306 generators. After removing equivalent TDI combinations, these generators are reduced to 11 inequivalent generators, as listed in Table \ref{table:5}.
\begin{widetext}
  \begin{table}[!ht]
    \centering
    \caption{Inequivalent generators of the first-generation TDI in 68 different generating sets.}\label{table:5}
      \renewcommand\arraystretch{1.5}
      \begin{tabular}{c|c}
        \toprule
        \toprule
        No. & generator\\
        \midrule
      1 & $(-D_1, -D_2, -D_3, D_1, D_2, D_3)$ \\
      2 & $(-1, -D_3, -D_1 D_3, 1, D_1 D_2, D_2)$ \\
      3 & $(0, 1 - D_2^2, D_1 - D_2 D_3, D_1 D_2 - D_3, D_2^2 - 1, 0)$ \\
      4 & $(0, D_1 - D_1 D_2^2, D_1^2 - 1, D_1^2 D_2 - D_2, 0, 1 - D_2^2)$ \\
      5 & $(0, D_2 - D_1 D_3, D_3 - D_1^2 D_3, 0, D_1^2 D_2 - D_2, D_1 D_2 - D_3)$ \\
      6 & $(0, 1-D_1 D_2 D_3, 0, D_1^2 D_3 - D_3, D_1^2 - 1, D_1 - D_2 D_3)$ \\
      7 & $(-D_2, D_1 - D_2 D_3 - D_1 D_2^2, D_1^2 - D_1 D_2 D_3 -1, D_1^2 D_2, D_1 D_2^2, 1)$ \\
      8 & $(D_2-D_1^2 D_2, D_1 + D_2 D_3 - D_1^2 D_2 D_3 - D_1 D_2^2, 0, D_1^3 D_3 + D_1^2 D_2 - D_1 D_3 - D_2, D_1^3 - D_1, D_1^2 - D_2^2)$ \\
      9 & $(1 - D_2^2, D_3 + D_1 D_2 - D_1 D_2^3 - D_2^2 D_3, D_3 - D_1 D_2 + D_1^2 D_2 - D_1 D_2^2 D_3, D_1^2 D_2^2 - 1, D_1 D_2^3 - D_1 D_2, 0)$ \\
      10 & $(-D_2^3 - D_1 D_2^2 D_3 + D_2 + D_1 D_3, D_2 D_3 - D_2^3 D_3, D_3^2 - D_2^2, D_1 D_2^2 D_3 + D_2 D_3^2 - D_2 - D_1 D_3, 0, D_2^2 D_3^2 - D_3^2)$ \\
      11 & $(D_2 - D_1^2 D_2, -D_1^2 D_2 D_3 + D_1 D_2^2 - D_1 + D_2 D_3, -D_1^3 D_2 D_3 - D_1^2 + D_1 D_2 D_3 + 1, 0, D_1^3 D_2^2 - D_1 D_2^2, D_1^2 D_2^2 - 1)$ \\
      \bottomrule
    \bottomrule
    \end{tabular}
  \end{table}  
\end{widetext}

For classical generators \eqref{eq:3-2-5} of the first-generation TDI, one has
\begin{equation}
	\begin{gathered}
		\mathcal{F}_{RI}(\alpha) =\beta,\; \mathcal{F}_{RI}(\beta) =\gamma,\; \mathcal{F}_{RI}(\gamma) =\alpha,\\
		\mathcal{F}_{RI}(\zeta) = \zeta,\; \mathcal{F}_{EI}(\zeta) = -\zeta.
	\end{gathered}
\end{equation}
It is showed that $\alpha$, $\beta$ and $\gamma$ are equivalent TDI combinations and $\zeta$ has a certain symmetry. Thus, there are only two TDI combinations that are not equivalent to each other in generating set \eqref{eq:3-2-5}.

For the modified first-generation TDI, as listed in Table \ref{table:3}, we obtain 148 distinct generating sets under 103680 different orderings, collectively comprising 1033 generators. After removing equivalent TDI combinations, these generators are reduced to only 5 inequivalent generators, as listed in Table \ref{table:6}.
\begin{widetext}
  \begin{table}[!ht]
    \centering
    \caption{Inequivalent generators of the modified first-generation TDI in 148 different generating sets.}\label{table:6}
      \renewcommand\arraystretch{1.5}
      \begin{tabular}{c|c}
      \toprule
      \toprule
      No. & generator\\
      \midrule
      1 & $(0, 1 - D_2 D_{2'}, D_1 - D_{2'} D_{3'}, D_1 D-2 - D_{3'}, D_2 D_{2'} - 1, 0)$ \\
      2 & $(0, 0, 1 - D_{1'} D_{2'} D_{3'}, D_2 - D_{1'} D_{3'}, D_{1'} D_2 D_{2'} - D_{1'}, D_2 D_{2'} - 1)$ \\
      3 & $(0, D_{1'} D_{3'} - D_2, D_1 D_{1'} D_{3'} - D_{3'}, 0, D_2 - D_1 D_{1'} D_2, D_3 - D_1 D_2)$ \\
      4 & $(0, D_{1'} - D_{1'} D_2 D_{2'}, D_1 D_{1'} - 1, D_1 D_{1'} D_2 - D_2, 0, 1 - D_2 D_{2'})$ \\
      5 & $(1 - D_2 D_{2'}, D_3 - D_2 D_{2'} D_3, D_1 D_3 - D_{2'}, D_1 D_2 D_3 - 1, 0, 0)$ \\ \bottomrule
      \bottomrule
    \end{tabular}
  \end{table}
\end{widetext}

For the generators of the modified first-generation TDI, one has
\begin{equation}\label{eq:5-16}
	\mathcal{F}_{RI}(g_2^{(1.5)}) =g_6^{(1.5)},\; \mathcal{F}_{RI}(g_4^{(1.5)}) = -g_3^{(1.5)}.
\end{equation}
which means that there are four mutually non-equivalent TDI combinations in generating set \eqref{eq:3-3-4}.

\section{Conclusions and discussion}\label{sec:6}
In this paper, we critically revisit the algebraic approach for time delay interferometry (TDI) and reveal an inability of any nonzero TDI combination to fully suppress laser frequency noise in typical space-based gravitational wave detectors. This necessitates alternative methodologies when pursuing second-generation TDI combinations in the context of time-varying armlengths. We have forged connections between TDI combinations of different generations and have put forth a systematic approach to identify higher-generation TDI combinations utilizing the generators of the lower-generation TDI. We also scrutinize a variety of zeroth-generation, first-generation, and modified first-generation TDI combinations under differing orderings and provide a list of non-equivalent generators for implementation in the search strategy.

The algebraic approach outlined in this study only involve time-delay operators, excluding the consideration of their inverses or time-advance operators. As a direction for future research, we intend to broaden the scope of the algebraic approach to incorporate time-advance operators, thereby merging algebraic techniques with geometric TDI and other research methods for a more comprehensive understanding of second-generation and subsequent higher-generation TDI combinations.

% \begin{widetext}
%   \begin{equation}
%   {\cal R}^{(\text{d})}=
%    g_{\sigma_2}^e
%    \left(
%      \frac{[\Gamma^Z(3,21)]_{\sigma_1}}{Q_{12}^2-M_W^2}
%     +\frac{[\Gamma^Z(13,2)]_{\sigma_1}}{Q_{13}^2-M_W^2}
%    \right)
%    + x_WQ_e
%    \left(
%      \frac{[\Gamma^\gamma(3,21)]_{\sigma_1}}{Q_{12}^2-M_W^2}
%     +\frac{[\Gamma^\gamma(13,2)]_{\sigma_1}}{Q_{13}^2-M_W^2}
%    \right)\;. 
%    \label{eq:wideeq}
%   \end{equation}
% \end{widetext}

\begin{acknowledgments}
  This work is supported by the National Key R$\&$D Program of China under Grants No.2022YFC2204602, No.2022YFC2204603, the Natural Science Foundation of China (Grants No.12247154, No.11925503), the Postdoctoral Science Foundation of China (Grant No.2022M711259).
\end{acknowledgments}

\appendix

\section{Proof of Proposition 2.1}\label{appendix:A}
In this appendix, we give the proof of the Proposition \ref{prop:1}. Recall that the homomorphism $\varphi: \mathcal{K}^4 \to \mathcal{K}^1$ is defined by
\begin{widetext}
  \begin{equation}\label{eq:appx-A-1}
    \varphi(D_1, D_2, D_3, D_{1'}, D_{2'}, D_{3'}) =q_3 (1 - D_2 D_3 D_1) + q_{1'} (D_3 D_1 - D_{2'}) +q_{2'} (D_1 - D_{3'} D_3 D_1) + q_{3'} (1 - D_{1'} D_1).
  \end{equation}
\end{widetext}
We prove that the kerner $\varphi^{-1}(0) = \{ (0, 0, 0, 0) \}$ by contradiction. Suppose that there is another solution
\begin{equation}
	(q_1, q_{1'}, q_{2'}, q_{3'}) \in \varphi^{-1}(0).
\end{equation}
Since $q_1$, $q_{1'}$, $q_{2'}$ and $q_{3'}$ are all noncommutative polynomials composed of variables $D_1, D_2, D_3, D_{1'}, D_{2'}, D_{3'}$, we investigate the monomials in these polynomials. There exist a longest monomial $m$ of the following form
\begin{equation}
	m = k D_{i_1} D_{i_2} \cdots D_{i_n},
\end{equation}
where $k \in K$ and $i_1, i_2, \cdots i_n \in \{ 1, 2, 3, 1', 2', 3' \}$. The length of the monomial $m$ is denoted by $n$, and the lengths of the other monomials are less than or equal to $n$. 

If $m$ is the monomial in $q_3$, then $- m D_2 D_3 D_1$ is the monomial of length $n + 3$, it must be cancelled by other monomial of length $n + 3$. According to our hypothesis, only $- q_{2'} D_{3'} D_3 D_1$ may have a monomial of length $n + 3$. Because of $D_2 D_3 D_1 \neq D_{3'} D_3 D_1$, the monomial $- m D_2 D_3 D_1$ can not be cancelled. Thus the longest monomial $m$ can not be in $q_3$.

If $m$ is the monomial in $q_{2'}$, then $- m D_{3'} D_3 D_1$ is the monomial of length $n + 3$. Based on the previous analysis, $q_1$ has no monomial of length $n$, the monomial $- m D_{3'} D_3 D_1$ can not be cancelled. Thus the longest monomial $m$ can not be in $q_{2'}$.

If $m$ is the monomial in $q_{3'}$, then $- m D_{1'} D_1$ is the monomial of length $n + 2$. $- q_3 D_2 D_3 D_1$ and $- q_{2'} D_{3'} D_3 D_1$ may have monomials of length $n + 2$. Because of $ D_{1'} D_1 \neq D_3 D_1$, the monomial $- m D_{1'} D_1$ can not be cancelled. Thus the longest monomial $m$ can not be in $q_{3'}$.

If $m$ is the monomial in $q_{1'}$, then $- m D_{2'}$ is the monomial of length $n + 1$. $- q_3 D_2 D_3 D_1$, $- q_{2'} D_{3'} D_3 D_1$ and $q_{3'} D_{1'} D_1$ may have monomials of length $n + 1$, but no monomial in these polynomials end with $D_{2'}$. The monomial $- m D_{1'} D_1$ can not be cancelled. Thus the longest monomial $m$ can not be in $q_{1'}$.

Through the above analysis, $m$ is not the monomial in $q_1$, $q_{1'}$, $q_{2'}$ or $q_{3'}$. This is a contradiction. Therefore, our assumption is wrong, and the kerner $\varphi^{-1}(0) = \{ (0, 0, 0, 0) \}$.

\section{Solutions of linear equations over a commutative polynomials ring}\label{appendix:B}

Consider a linear equation
\begin{equation}\label{eq:appx-B-1}
	f_1 q_1 + f_2 q_2 + \cdots + f_m q_m = 0,
\end{equation}
where variables $q_i$ and coefficients $f_i$ $(i = 1, 2, \ldots, m)$ are evaluated in the polynomial ring $K[x_1, \ldots, x_n]$.

Let
\begin{equation}\label{eq:appx-B-2}
	\mathbf{f} = (f_1, f_2, \ldots, f_m)^T,\; \mathbf{q} = (q_1, q_2, \ldots, q_m)^T,
\end{equation}
then Eq. \eqref{eq:appx-B-1} becomes
\begin{equation}\label{eq:appx-B-3}
	\mathbf{f}.\mathbf{q} = 0.
\end{equation}

Let the set of solutions of Eq. \eqref{eq:appx-B-3} be $S_q$. Because that the polynomial ring $K[x_1, \ldots, x_n]$ is a Noetherian ring, $S_q$ is finitely generated. Let
\begin{equation}\label{eq:appx-B-4}
	\langle \mathbf{f} \rangle = \langle f_1, f_2, \ldots, f_m \rangle
\end{equation}
be the ideal generated by the polynomials $f_i (i = 1, 2, \ldots, m)$. The Gröbner basis \cite{Ralf1996An} $g_i (i = 1, 2, \ldots, l)$ of the ideal $\langle \mathbf{f} \rangle$ can be obtained by specifying a monomial ordering. One has
\begin{equation}\label{eq:appx-B-5}
	\langle \mathbf{f} \rangle = \langle \mathbf{g} \rangle.
\end{equation}

Let $C_1$ be an $l \times m$ matrix and let $C_2$ be an $m \times l$ matrix such that
\begin{align}
	\mathbf{g} &= C_1 \mathbf{f}, \label{eq:appx-B-6} \\
	\mathbf{f} &= C_2 \mathbf{g}. \label{eq:appx-B-7}
\end{align}

Substituting Eq. \eqref{eq:appx-B-6} in Eq. \eqref{eq:appx-B-7}, one has
\begin{equation}\label{eq:appx-B-8}
	\mathbf{f} = C_2 \mathbf{g} = C_2 C_1 \mathbf{f},
\end{equation}
which implies
\begin{equation}\label{eq:appx-B-9}
	(I - C_2 C_1) \mathbf{f} = \mathbf{0}.
\end{equation}

Let
\begin{equation}\label{eq:appx-B-10}
	I - C_2 C_1 = (s_1, s_2, \ldots, s_m)^T.
\end{equation}
Then $s_i (i = 1, 2, \ldots, m)$ are solutions of Eq. \eqref{eq:appx-B-2}.

If the polynomial ring $K[x_1, \ldots, x_n]$ have only one variables $x_1$, i.e., $n=1$, one has
\begin{equation}\label{eq:appx-B-11}
	\langle s_1, s_2, \ldots, s_m \rangle = S_q.
\end{equation}
This is the situation for the zeroth-generation TDI solutions, and we can get the generators of Eq. \eqref{eq:3-1-5}.

If the polynomial ring $K[x_1, \ldots, x_n]$ have only two or more variables, i.e., $n \geq 2$, one has
\begin{equation}\label{eq:appx-B-12}
	\langle s_1, s_2, \ldots, s_m \rangle \subset S_q.
\end{equation}
This is the situation for the first-generation TDI solutions and the modified-generation TDI solutions. Thus, we need to search for more generators for Eq. \eqref{eq:3-2-4} or \eqref{eq:3-3-3}.

Denote the nonzero S-polynomials \cite{Ralf1996An} of the polynomials $s_i (i = 1, 2, \ldots, m)$ by $sp_i (i = 1, 2, \ldots, m_2)$. According to the calculation formula of SC-polynomials, we can get an $m_2 \times l$ matrix $A_1$ such that
\begin{equation}\label{eq:appx-B-13}
	A_1 \mathbf{g} = (sp_1, sp_2, \ldots, sp_l)^T,
\end{equation}
where $m_2 \leq m(m-1)/2$ and each row of $A_1$ has only two nonzero entries.

Since $g_i (i = 1, 2, \ldots, l)$ is a Gröbner basis, by using polynomial reduction, we can find another $m_2 \times l$ matrix $A_2$ such that
\begin{equation}\label{eq:appx-B-14}
	A_2 \mathbf{g} = (sp_1, sp_2, \ldots, sp_l)^T.
\end{equation}
Substracting Eq. \eqref{eq:appx-B-13} from Eq. \eqref{eq:appx-B-14} gives
\begin{equation}\label{eq:appx-B-15}
	(A_1 - A_2) \mathbf{g} = \mathbf{0}.
\end{equation}
Substituting Eq. \eqref{eq:appx-B-6} into Eq. \eqref{eq:appx-B-15} gives
\begin{equation}\label{eq:appx-B-16}
	(A_1 - A_2) \mathbf{g} = (A_1 - A_2) C_1 \mathbf{f} = \mathbf{0}.
\end{equation}

Let
\begin{equation}\label{eq:appx-B-17}
	(A_1 - A_2) C_1 = (\bar{s}_1, \bar{s}_2, \ldots, \bar{s}_{m_2})^T.
\end{equation}
Then $\bar{s}_i (i = 1, 2, \ldots, l)$ are also solutions of Eq. \eqref{eq:appx-B-2} and we have
\begin{equation}\label{eq:appx-B-18}
	\langle s_1, s_2, \ldots, s_m, \bar{s}_1, \bar{s}_2, \ldots, \bar{s}_{m_2} \rangle = S_q.
\end{equation}

The number of the resulting generators is
\begin{equation}
	m + m_2 \leq \frac{m(m+1)}{2}.
\end{equation}
To simplify the result, we can reduce the number of the generators. For a fixed ordering, a unique reduced Gröbner basis can be obtained \cite{Ralf1996An}, which form a generating set of $S_q$. In the next appendix, we describe the various orderings encountered in solving the Eq. \eqref{eq:appx-B-1}.

\section{Description for the orderings about polynomials}\label{appendix:C}
When solving algebraic equations for the first-generation TDI or the modified first-generation TDI, we encounter different types of orderings and the resulting generators depend on the choice of the four orderings~\cite{Ralf1996An}:
\begin{enumerate}
	\item Variable ordering. There are $n!$ different variable orders for $n$ variables, for example,
	\begin{equation}
		x_1 \succ x_2 \succ \cdots \succ x_n.
	\end{equation}
	The variable order determines the lexicographical order $\succ_\text{lex}$ in the monomial order.
	\item Monomial ordering. In the case of a single variable, there is only one monomial order. For the case of multiple variables, there are infinite kinds of monomial orderings, such as the lexicographical ordering $\succ_\text{lex}$, the degree lexicographic ordering $\succ_\text{deglex}$ (or the graded lexicographic ordering) and the degree reverse lexicographic ordering $\succ_\text{degrevlex}$(or the graded reverse lexicographic ordering). Specifically, the degree lexicographic ordering compares the monomials by their total degree and then uses the lexicographic ordering, and the degree reverse lexicographic ordering compares the monomials by their total degree and then uses the reverse lexicographic ordering. The monomial ordering can determine the leading term of a polynomials.
	\item Position ordering. For a polynomial vector
	\begin{equation}
		\mathbf{q} = (q_1, q_2, \ldots, q_m) = q_1 \mathbf{e}_1 + q_2 \mathbf{e}_2 + \cdots + q_m \mathbf{e}_m,
	\end{equation}
	where $\mathbf{e}_i$ are the standard basis
	\begin{equation}
		\mathbf{e}_1 = (1, 0, \ldots, 0),\; \mathbf{e}_2 = (0, 1, \ldots, 0),\; \ldots,\; \mathbf{e}_m = (0, 0, \ldots, 1),
	\end{equation}
	the position ordering gives an order of the standard basis, such as $\mathbf{e}_1 \succ \mathbf{e}_2 \succ \cdots \succ \mathbf{e}_m$. For the vector $\mathbf{q}$ with $m$ components, there are $m!$ different position orderings.
	\item Relative ordering of terms and positions. There are usually two types of this ordering. If we compare the polynomials first and then compare the positions of the polynomials, this is called "term over position" (TOP). If we compare the positions of the polynomials first and then compare the polynomials, this is called "position over term" (POT).
\end{enumerate}

Combining with the above orderings, a reasonable ordering for the terms of polynomial vectors with $n$ components over a polynomial ring $K[x_1, \ldots, x_m]$ can be defined and used to obtain the unique reduced Gröbner basis. See monographs \cite{Ralf1996An,Cox2015Ideals} for specific definitions and examples.

For the zeroth-generation TDI and Eq. \eqref{eq:3-1-4}, $n=1$ and $m=4$. The numbers of the orderings considered in the paper is $n! \times 1 \times m! \times 2 = 48$.

For the first-generation TDI and Eq. \eqref{eq:3-2-4}, $n=3$ and $m=4$. The numbers of the orderings considered in the paper is $n! \times 3 \times m! \times 2 = 864$.

For the modified first-generation TDI and Eq. \eqref{eq:3-2-4}, $n=6$ and $m=4$. The numbers of the orderings considered in the paper is $n! \times 3 \times m! \times 2 = 103680$.

% The \nocite command causes all entries in a bibliography to be printed out
% whether or not they are actually referenced in the text. This is appropriate
% for the sample file to show the different styles of references, but authors
% most likely will not want to use it.
% \nocite{*}

\bibliographystyle{h-physrev}
\bibliography{apsref}% Produces the bibliography via BibTeX.

\end{document}